%% file: pop.tex
\documentclass[revtex4]{emulateapj}
\usepackage{apjfonts}
\usepackage{graphicx}
\usepackage{amsmath}
\usepackage{amssymb}
\usepackage{amsfonts}
\usepackage{lscape}
\usepackage{longtable}
\usepackage{rotating}
\shorttitle{X-ray binaries in Cen~A}
\shortauthors{Burke et al.}

\begin{document}

\title{Spectral Properties of X-ray Binaries in Centaurus A}

\author{Mark~J.~Burke\altaffilmark{1,2}, 
Somak~Raychaudhury\altaffilmark{1,15}, 
Ralph~P.~Kraft\altaffilmark{2}, 
Thomas~J.~Maccarone\altaffilmark{9}, 
Nicola~J.~Brassington\altaffilmark{3},
Martin~J.~Hardcastle\altaffilmark{3},
Jouni~Kainulainen~\altaffilmark{7},
Kristin~A.~Woodley\altaffilmark{5,16},
Joanna~L.~Goodger\altaffilmark{3},
Gregory~R.~Sivakoff\altaffilmark{4}, 
William~R.~Forman\altaffilmark{2}, 
Christine~Jones\altaffilmark{2},
Stephen~S.~Murray\altaffilmark{5,2},
Mark~Birkinshaw\altaffilmark{8,2},
Judith~H.~Croston\altaffilmark{9},
Daniel~A.~Evans\altaffilmark{2},
Marat~Gilfanov\altaffilmark{10,14},
Andr\'{e}s~Jord\'{a}n\altaffilmark{2,11},
Craig~L.~Sarazin\altaffilmark{12},
Rasmus~Voss\altaffilmark{13},
Diana~M.~Worrall\altaffilmark{8,2}, and
Zhongli~Zhang\altaffilmark{10}
}

\altaffiltext{1}{School of Physics and Astronomy, University of
  Birmingham, Edgbaston, Birmingham, B15 2TT, UK 
\email{mburke@star.sr.bham.ac.uk}}

\altaffiltext{2}{
  Harvard-Smithsonian Center for Astrophysics, 60 Garden Street,
  Cambridge, MA 02138, USA}

\altaffiltext{3}{School of Physics, Astronomy, and Mathematics, 
			University of Hertfordshire,
			Hatfield, AL10 9AB, UK}

\altaffiltext{4}{Department of Physics, University of Alberta,
Edmonton, Alberta T6G 2E1, Canada}

\altaffiltext{5}{Department of Physics and Astronomy, University of British Columbia, Vancouver BC V6T 1Z1, Canada}

\altaffiltext{6}{Department of Physics and Astronomy, Johns Hopkins University, 3400 N. Charles Street, Baltimore, MD 21218, USA}

\altaffiltext{7}{Max-Planck-Institute for Astronomy, K\"{o}nigstuhl 17, 69117 Heidelberg,Germany}

\altaffiltext{8}{HH Wills Physics Laboratory, University of Bristol, Tyndall Avenue, Bristol BS8 1TL, UK}

\altaffiltext{9}{School of Physics and Astronomy, University of Southampton, Southampton, SO17 1BJ, UK}

\altaffiltext{10}{Max Planck Institut f\"{u}r Astrophysik, Karl-Schwarzschild-Str. 1, D-85741, Garching, Germany}

\altaffiltext{11}{Departamento de Astronom\'{i}a y Astrof\'{i}sica, Pontificia Universidad Cat\'{o}lica de Chile, Casilla 306, Santiago 22, Chile}

\altaffiltext{12}{Department of Astronomy, University of Virginia, P.O. Box 400325, Charlottesville, VA 22904-4325, USA}

\altaffiltext{13}{Department of Astrophysics/IMAPP, Radboud, University Nijmegen, PO Box 9010, NL-6500 GL Nijmegen, the Netherlands.}

\altaffiltext{14}{Space Research Institute, Russian Academy of Sciences, Profsoyuznaya 84/32, 117997 Moscow, Russia}

\altaffiltext{15}{Department of Physics, Presidency University, Kolkata 700 073, India}

\altaffiltext{16}{UC Santa Cruz, University of California Observatories, 1156 High Street, Santa Cruz, CA 95064, USA}

\begin{abstract}
We present a spectral investigation of X-ray binaries in NGC~5128 (Cen A), using six 100 ks \emph{Chandra} observations taken over two months in 2007.  We divide our sample into thermally and non-thermally dominated states based on the behavior of the fitted absorption column $N_H$, and present the spectral parameters of sources with $L_x\gtrsim2\times10^{37}~{\rm erg~s^{-1}}$.  The majority of sources are consistent with being neutron star low mass X-ray binaries (NS LMXBs) and we identify three transient black hole (BH) LMXB candidates coincident with the dust lane, which is the remnant of a small late-type galaxy.  Our results also provide tentative support for the apparent `gap' in the mass distribution of compact objects between $\sim2-5~M_\odot$.

We propose that BH LMXBs are preferentially found in the dust lane, and suggest this is because of the younger stellar population.  The majority ($\sim70-80\%$) of potential Roche-lobe filling donors in the Cen A halo are $\gtrsim12$ Gyr old, while BH LMXBs require donors $\gtrsim1 {\rm M_\odot}$ to produce the observed peak luminosities. This requirement for more massive donors may also explain recent results that claim a steepening of the X-ray luminosity function with age at $L_x \geq 5\times 10^{38}~{\rm erg~s^{-1}}$ for the XB population of early-type galaxies; for older stellar populations, there are fewer stars $\gtrsim1 {\rm M_\odot}$, which are required to form the more luminous sources.

\end{abstract}


\section{Introduction}
\label{s:intro}
Population studies of extragalactic X-ray binaries (XBs) located beyond the Local Group have been made possible thanks to the excellent sensitivity and spatial resolution of the \emph{Chandra} X-ray observatory \citep{2000SPIE.4012....2W}.  For over a decade, astronomers have resolved the hard X-ray spectral component of galaxies at sub-arcsecond precision into predominantly non-nuclear point sources \citep{2000ApJ...544L.101S}. Much research has focused
on associations with globular clusters, which contain the majority of X-ray point sources in some elliptical galaxies \citep[e.g.][]{2001ApJ...557L..35A}.  It has been observed that metal rich clusters are clearly favored as the hosts of X-ray sources \citep{2002ApJ...574L...5K} and that the densest clusters preferentially host X-ray sources \citep{2007ApJ...671L.117J}.
Ultraluminous X-ray sources, which display isotropic luminosities $L_x>10^{39}~{\rm erg~s^{-1}}$ \citep{1989ARA&A..27...87F}, have been studied in great detail by both \emph{Chandra} and  \emph{XMM-Newton}. Optical observations have been used to detect counterparts for these sources in star-forming galaxies \citep{2008MNRAS.387...73R}, and to demonstrate association with globular clusters in a few early-type galaxies \citep[e.g.][]{2007Natur.445..183M,2012arXiv1206.5304B}.

The X-ray luminosity function (XLF) is a well-studied statistical characteristic of the X-ray source population for a given galaxy. \cite{2002A&A...391..923G} conducted a study of Galactic XBs and found that the XLF of high-mass X-ray binaries (HMXBs) tightly follows a power law above $10^{37}~{\rm erg~s^{-1}}$, while that of low-mass X-ray binaries (LMXBs) experiences a sharp cut-off above a few $10^{38}~{\rm erg~s^{-1}}$.  Usually there are too few counts to derive the source spectra directly and XLFs are produced using an assumed spectrum, typically a power law with $\Gamma\sim1.5-1.7$, which approximates the emission from Galactic XBs.  An LMXB population dominates the XLF of discrete sources in early-type galaxies \citep[see~][~for review]{2006ARA&A..44..323F}, due to their older stellar population; early-type galaxies contain relatively few  high luminosity($> 10^{38} {\rm \, erg \, s}^{-1}$) sources compared to late-type star-forming galaxies.

The tendancy of LMXBs to dominate the discrete source XLFs in early-type galaxies can be linked to the study of XB populations in late-type galaxies, where two distinct populations have been observed \citep{2009ApJ...705.1632P}. One population, associated primarily with the bulge, follows an XLF that steepens after a few $\times10^{38}~{\rm erg~s^{-1}}$, while the other population is associated with the spiral arms and has an XLF consistent with that of HMXBs.  The latter population has both softer colours and higher luminosities than Galactic, wind-driven HMXBs, most of which are accretion-powered pulsars with high magnetic fields, and the spectra are well-described by relatively cool disk blackbodies (0.1-1.1 keV), suggesting that they too are  Roche-lobe-filling accretion driven systems.

A luminosity break at $L_x~\sim 2.5-5.0  \times 10^{38}~{\rm  erg~s^{-1}}$ has been seen in the discrete source population of numerous early-type galaxies \citep[e.g.,][]{2001ApJ...556..533S} and is potentially the Eddington-limited extent of the brightest neutron star (NS) LMXBs.  The brightest sources are super-Eddington for a 1.4 ${\rm  M_\odot}$ NS and are probably BH XRBs or the more massive extreme of the NS LMXB population.  The luminosity break was examined in detail over a sample of nearby elliptical galaxies  by \cite{2010ApJ...721.1523K}, who found that the lack of sources with $L_x >5\times 10^{38}~{\rm  erg~s^{-1}}$ was more pronounced for older ($>5$ Gyr) galaxies.  The observed steepening of the break in the XLF suggests that young early-type XB populations are intermediate in nature between those of star-forming galaxies and old early-types.


NGC~5128 (Centaurus A) is the nearest optically luminous early-type galaxy,  at a distance of 3.7 Mpc \citep{2007ApJ...654..186F}, with $M_B = -21.1$. A small late-type galaxy is currently merging  with Cen A \citep{1979ApJ...232...60G}; however, the galaxies remain poorly mixed \citep{2006ApJ...645.1092Q}.  The central regions of the galaxy are notable for the presence of vast dust lanes that contain many active star forming regions.  \cite{2011A&A...526A.123R} compared simulated color-magnitude diagrams with deep ACS/HST photometry and concluded that \emph{at least} $70\%$ of stars in NGC~5128 formed $12 \pm 1$ Gyr ago and a smaller population of more metal rich stars formed in the last 2-4 Gyr. Therefore we expect the population of XBs to be dominated by LMXBs that possess old main sequence companions of $M < 1 {\rm M_\odot}$ with the potential for a small component from LMXBs with more massive companions.

Six 100~ks \emph{ Chandra} observations of NGC~5128 were taken as part of the Cen~A Very Large Project (VLP) spanning the course of 2 months in 2007 \citep{2007ApJ...671L.117J}.  These observations led to investigation of the source's striking X-ray features such as the AGN jet (\cite{2007ApJ...670L..81H}, \cite{2008ApJ...673L.135W}, \cite{2010ApJ...708..675G}), radio-lobe shock \citep{2009MNRAS.395.1999C} and the extended gaseous emission \citep{2008ApJ...677L..97K}.  The relationship between XBs and globular clusters was investigated by \cite{2009ApJ...701..471V}, who confirmed the presence of a low-luminosity break in the XLF at $L_x \sim 1.5-4 \times  10^{37}~ {\rm erg~s^{-1}}$ and a lack of GC sources with $L_x < 3 \times 10^{36}~ {\rm erg~s^{-1}}$. \cite{2009ApJ...701..471V} suggested that this dearth of faint sources may indicate that GC LMXB companions are He rich, which possess a larger critical mass accretion rate $\dot{M}_{crit}$ to become transients.  Sources approaching the corresponding luminosity to $\dot{M}_{crit}$ become unstable, and are therefore always above or below this luminosity, leaving a gap in the XLF. However, it now appears more likely that the low-luminosity break results from a change in the disk instability criterion at low accretion rates as the disk spectrum peaks at longer wavelengths \citep{2012A&A...537A.104V}. These Cen~A data were also used by \cite{2011A&A...533A..33Z} as part of a large sample of GC-LMXBs from many elliptical galaxies, and discrepancies between the GC and field XLFs were found across $L_x\sim 10^{36}-10^{39} {\rm erg~s^{-1}}$.

The current era of deep X-ray observations has enabled more detailed study of individual XBs beyond the Local Group.  \cite{2010ApJ...725.1805B} and \cite{2010ApJ...725.1824F} present the results of spectral fitting the brightest sources in the early-type galaxies NGC~3379 and NGC~4278, down to a limiting unabsorbed luminosity of $\sim 1.2 \times 10^{38}~ {\rm erg~s^{-1}}$.  These samples contained 8 and 7 XBs, respectively, of which 7 were coincident with globular clusters (GCs). These works adopted a diagnostic approach to spectral fitting, inferring the state of a source based on fitting simple spectral models,  a multi-colored disk blackbody and power law. The true state of the source can be inferred from the behavior of the absorption parameter $N_H$, which simulations showed to behave in a characteristic way depending on the true state of the source.  This method essentially assesses the relative contribution of thermal and non-thermal emission to the spectrum, and is fully described by \cite{2010ApJ...725.1805B}~(see section~\ref{s:res}).  A further study of 18 transient-type sources from galaxies NGC~3379, NGC~4278 and NGC~4697 by \cite{2012arXiv1206.5304B} found a host of exotic sources, including a GC ULX in outflow and an unusually luminous bursting source.  They distinguished spectral states down to $\sim 10^{38}~{\rm erg~s^{-1}}$.

In Cen~A we expect to detect both neutron star (NS) and black hole (BH) XBs, both of which can be transients.  It is generally thought that the transient behavior occurs due to the disk ionization instability mechanism (DIM), which was first developed for explaining the dwarf nova outbursts of cataclysmic variables \citep{1984PASP...96....5S}.  The essential feature of this model is that the viscosity of ionized gas is larger than that of neutral gas. Its application to X-ray binaries is complicated by irradiation of the outer accretion disk, which is a considerably more important factor \citep{2001A&A...373..251D}.  Recently, \cite{2012MNRAS.424.1991C} tested the DIM with a  population of Galactic sources, finding that the critical mass accretion rate above which sources are persistent is lower than that predicted when irradiation is not taken into account.  Black holes are more massive than neutron stars, but have similar radiative efficiencies, and a consequence is that the outer disk temperatures for black hole X-ray binaries at a given orbital period will be smaller than those for neutron star X-ray binaries; thus black hole systems are far more likely to be transient than are neutron star systems \citep{1996ApJ...464L.127K}.  

An empirical understanding of the properties of transient X-ray binaries has started to develop.  The size of the accretion disk should determine the peak outburst luminosity \citep[e.g.][]{1998MNRAS.301..382S}, and this has been borne out as large samples of such transient outbursts have developed (\cite{2004MNRAS.355..413P}; \cite{2010ApJ...718..620W}).  As sources change in luminosity, they follow loops in a hardness-intensity diagram \citep{2003A&A...399.1151M}, indicating that they are changing spectral shapes hysteretically. In general, the spectral changes occur rapidly, with X-ray binaries in outburst spending most of their time in just a few spectral states.  \cite{2006ARA&A..44...49R} posit that BH LMXBs possess spectra characterized by three key spectral states.  The thermal-dominant state, where the emission appears to be dominated by a $\sim 1$ keV  multicolor disk blackbody, is essentially the same as the standard geometrically thin, optically thick accretion disk of \cite{1973A&A....24..337S}. During the ingress and egress of outburst, the source experiences a hard power-law state of $\Gamma \sim 1.7$, the emission likely due to inverse Compton scattering in an optically thin, geometrically thick region \citep{1975ApJ...195L.101T}.  Near the peak of outburst, some sources also exhibit a steep power law state $\Gamma\sim2.5$ extending to MeV energies with a significant thermal component also present.  These spectral states are also associated with the rapid variability of the sources \citep{2001ApJS..132..377H} and changes in the radio jet properties \citep{2004MNRAS.355.1105F}. 

NS LMXBs generally show similar spectral state phenomenology to BH systems \citep{1994ApJS...92..511V}, but have some differences since the NS surface provides a boundary layer.   Nonetheless, in both cases, low/hard type spectra are typically seen below $\sim 2\%$ of the Eddington luminosity, except during the hysteretic intervals near the beginning of outbursts \citep{2003A&A...409..697M}, and steep power law states are generally seen only at very high luminosities, near the Eddington limit \citep{2006ARA&A..44...49R}.  This spectral state phenomenology allows, with high quality spectra, a source to be classified as a candidate BH XB on the basis of its having a cool accretion disk at ~$10^{38}~{\rm erg~s^{-1}}$, as discussed by \cite{1984ApJ...281..354W}. \cite{2012ApJ...749..112B} used the Cen~A VLP data to show that such distinctions are now possible for sources outside the Local Group, presenting evidence that CXOU J132527.6-430023 (S14, Table 1) is a BH LMXB.

The relative proximity of Cen~A, coupled with the superb quality of these data, allows us unrivaled insight into the XBs of an early-type galaxy.  In this work we divide sources into thermal and non-thermally dominant states, where appropriate, down to a luminosity of $2\times 10^{37}~{\rm erg~s^{-1}}$ -- reliably measuring the spectral properties of XBs at similar luminosities to those found in the Local Group.

\section{Data Preparation}
\label{s:obs}
\subsection{Source Detection and Alignment}
Each of the six 100 ks observations was analysed using CIAO 4.3, and was reprocessed using the \emph{ chandra\_repro} script. The \emph{ destreak} tool was used to remove the ACIS readout streak, caused by the bright Cen~A nucleus. Light curves of each event file were produced using \emph{ dmextract} to check for background flares, which were not present.
To search for point sources, we used a 0.5-2.0 keV event file for each observation. We used this band because the central AGN is so bright (6-10 count ${\rm s^{-1} }$ in ACIS-I) that the wings of the PSF contain a significant number of counts, the PSF being broader at higher energies for \emph{Chandra}. An exposure map was created for each file, weighted by the typical power law spectrum of an LMXB, with a photon index of $\Gamma = 1.7$ and absorption column at the Galactic value of $N_H=8.4 \times 10^{20}~{\rm cm^{-2}}$ \citep{1990ARA&A..28..215D}. Inside a 5$\arcmin$ region centered on the Cen~A nucleus, we located point sources using the CIAO tool \emph{wavdetect} using the spectrally weighted exposure map, wavelet scales of $1.0$ to $16.0$ in steps of $\sqrt{2}$, a threshold significance of $10^{-6}$ and a maximum of $\sim1$ false source per ACIS chip.  All subsequent work is within 5$\arcmin$ of the Cen~A nucleus, which corresponds to $\sim $ the half-light radius of Cen A \citep{1979ApJ...232...60G}.

The six observations were aligned by applying an appropriate x-y shift to five of the aspect solution files using the CIAO tool \emph{reproject\_events}. All observations were aligned to the point source positions from obsID 8490, chosen because of the proximity of the Cen~A nucleus to the ACIS-I focus. Each shift was calculated from the mean offset in $\alpha$ and $\delta$ obtained performing $2\arcsec$ matching between point source lists from the two observations.   To reduce the effect of false matches on our offset correction, we found the mean offset in $\alpha$ and $\delta$ and then calcuated the mean offset within $\pm0.5\arcsec$ of this mean. By applying $5\arcsec$ shifts in $\alpha$ and $\delta$, subsequent $2\arcsec$ matching found $\sim 6$ false matches between source lists.

To allow for accurate analysis of globular cluster (GC) sources in our subsequent work, we utilized the well known LMXB-GC connection to align our observations to GC positions \citep{2012AJ....143...84H}, again using $2\arcsec$ matching (we estimate an approximate GC size of $\sim 2\arcsec$), we calculated the mean x-y shift between LMXB positions in obsID 8490 and the globular clusters, and applied this shift to all obsID, maintaining the initial alignment of the X-ray data.  

We created a merged event file using the CIAO script \emph{merge\_all} and a corresponding exposure map, as before.  The point source list produced by a subsequent run of \emph{wavdetect} was used as a master list of source positions that was consistent with the positions of the individual runs on each obsID, for each observation in which the source was detected.  Circular extraction regions were produced, centered on the chip position of each source, whether it is detected in an observation or not.  As was the case for the work of \cite{2012ApJ...749..112B}, these regions had radius $r_{p7}$, equal to $90\%$ of the 7 keV extraction radius at that chip position.  We then excluded regions in the jet, radio lobe and nucleus from our source list.  Additional region files were produced, covering the removed read-out streak of the bright central AGN.  Background region files were then created based on annuli from $2r_{p7}$ to $4r_{p7}$, with all source regions and the read-out region excluded.  Subsequently we tested for source confusion, which was a significant problem in observations 7798 and 7799, where the point source population is further off-axis and the point-spread function is much wider.  Sources found to have a neighbor at angular half-distance $d/2$ inside $r_{p7}$ were given a new extraction region radius of $d/2$ provided that $d/2 \geq r_{p2}$ (where $r_{p2}$ is 90\% PSF radius at 2keV), while sources with $d/2 < r_{p2}$ were declared confused (Table~\ref{tab:src}).  Spectral fitting using various sized extraction regions for isolated, off-axis sources determined that the normalisation found from fitting was consistent between $r_{p2}$ and $r_{p7}$, but increased systematically as the extraction radius decreased further.

The net counts inside each extraction region were estimated using \emph{dmextract} to the requisite properties of the source and background region, followed by using \emph{aprates} to calculate a $90\%$ confidence bound for each value or, where appropriate, a $90\%$ confidence upper-limit.  Sources were then sorted, based on their highest count-flux observed in any individual observation and names assigned based on this ranking.  We present these results in Table ~\ref{tab:src}, and emphasize that these are \emph{not} the estimated source counts, but the net source counts present in a given extraction region.  This indicates the quality of the resulting spectra we extracted for a given source.  

\subsection{Source Selection}
\label{sec:select}
The presence of short-term variability may indicate important spectral changes within an X-ray source, and so we choose to exclude sources that vary during an observation from our sample.  To assess the variability we made use of the Gregory-Loredo algorithm \citep{1992ApJ...398..146G}, implemented as the CIAO tool \emph{glvary}\footnote{$http://cxc.harvard.edu/ciao/ahelp/glvary.html$}.  This produces an optimally-binned lightcurve for each source in each observation, and calculates an odds ratio of each lightcurve against a constant count-rate, with a ratio of 10 indicating the highest probability of variation and $0$ being consistent with a constant count rate.  A source is defined as showing definite variability when this index is $\geq 6$.  The maximum ratio for each source is reported in Table ~\ref{tab:src}. Given the potential for rapidly changing spectra within an observation, sources with a maximum score of 6 or higher were removed from our sample for spectral analysis.    

At this point we also flagged sources within $20\arcsec $ of the Cen~A nucleus, which require more careful extraction to deal with the significant flux contribution present from the nucleus.  This contribution likely exceeds the source counts in some observations for any sources with fewer than $\sim300$ counts, depending on how far off-axis the nucleus is in the observation.  Analysis of the sources that showed definite intra-observation variability and the sources within $20\arcsec $ of the nucleus will be reported in a future work. 

Table~\ref{tab:src} summarises some basic source data; the celestial position, net counts in each source region for each observation and the intra-observational variability ($\rm {G-L_{max}}$).  We also include our eventual source classification of persistent (P) or transient (T) black hole candidates (BHC), neutron star candidates (NSC), foreground stars (FG), active galactic nuclei (AGN) and also denote S48 as a highly-magnetised NSC ($\beta$).


The distribution of hardness ratios showed that source S50 was far softer than the rest of the population, with S/H$\sim 9 $ where S is the counts from 0.5--2.0 keV and H is the counts from 2.0--8.0 keV. The spectrum of this source was well fit with an absorbed apec model \citep{1977ApJS...35..419R}, peaking at 0.9 keV, across all observations, consistent with expectations from a foreground star. Inspection of the data taken with the Inamori Magellan Areal Camera and Spectrograph (IMACS) camera on the Magellan Observatory Baade telescope \citep{2012AJ....143...84H}, showed that S50 is coincident with an object consistent with being stellar.  We used ishape, which determines an object's shape parameters by analytically convolving a \emph{King62} \citep{1962AJ.....67..471K} model with the PSF of the image, which is then subtracted from the input image of the object itself.  From the residual image, the pixels are assigned a weighting based on their deviation from other pixels at the same distance from the center of the object, and then a reduced $\chi^2$ is calculated.  The initial parameters of the model are adjusted and the process is repeated with the new model until a minimation of the $\chi^2$ is obtained and convergence is reached.   Our best fit with \emph{ishape} indicates that our object has a FWHM of 0.02 pc, consistent with a star, and very different from the typical globular cluster which are generally 2-4 pc in size \citep{2010MNRAS.401.1965H}, and we do not consider it further.

\input{sourcedata}

\section{Spectral Analysis}
\label{sec:multi}
Our primary objective is using spectral fitting of sample models to infer the spectral state of each source.  Ideally, this would involve fitting all observations of a given source together, taking advantage of more bins to reduce the high-likelihood region of the parameter space. A more constrained  absorption parameter, $N_H$ provides a stronger argument for a given spectral state (see section~\ref{s:res}).  However, it is often the case with joint fitting that spectra from different states of a source will produce a best fit that, while well constrained and even statistically `acceptable', may poorly represent the true spectra individually.  To address this issue, we designed a comparative test between the spectra to allow us to judge the extent of spectral changes for each source, and thus the appropriateness of joint fits.  Since we allow for changes in luminosity for a given spectral state, and only sought to determine a difference in the spectral shape of a source, the normalization of each spectrum is kept free in the joint fits. 

Spectra were extracted using the CIAO script \emph{specextract} for each observation that had  $> 100$ net counts.  We chose to fit only spectra where we had either $>150$ counts in one observation, or more than two observations with $>100$ counts.  


\subsection{Spectral Variation}
\label{sec:specvar}
Using the spectral fitting package \emph{XSPEC} \citep{1996ASPC..101...17A}, we performed spectral fits using an absorbed power law to the ungrouped spectrum, then used the covariance matrix of each fit in a subsequent Markov chain Monte Carlo (MCMC) exploration of the parameter space for each spectrum of each source in the sample. 

\begin{figure}[htb!]\center
{\includegraphics[width=1.05\hsize, angle=270 ,trim=0cm 0cm 0.5cm 0cm, clip=true]{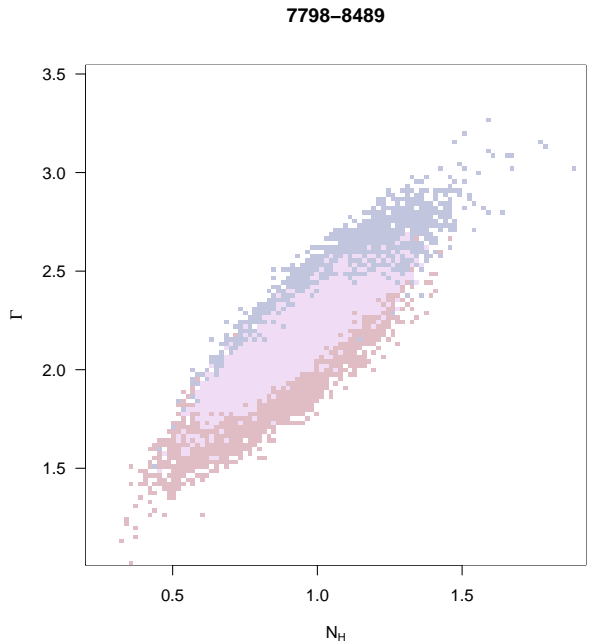}
\caption{Example MCMC output binned over $N_H$ and $\Gamma$ for two spectra of S24.  We show 90\% confidence regions, with the overlap indicated by the lighter region.  There is an $\sim 75\%$ probability that true combinations of $N_H$ and $\Gamma$ are shared between the two MCMCs. \label{fig:MCMC}}}
\end{figure}

We then tried to compare each spectrum with the (up to) five other spectra for a given source.  For the MCMC results of a particular spectrum, we binned the parameter values of $N_H$ and $\Gamma$, which describe the spectrum, into a $100\times 100$ grid; the resulting bin widths were used to extend the grid, if necessary, to cover the other set of parameter co-ordinates.  This approach was intended to prevent the binning being too coarse or fine, the extent of the MCMC-explored region of the parameter space increasing as the number of counts in a given spectrum decreased.  The number of counts in each bin, normalized by the total number of samples, gives the approximate 2D posterior probability density for the $N_H$ -- $\Gamma$ space.  The probability that the true coordinate is in an overlapping bin is the product of the sum of the probabilities for all the overlapping bins from both distributions.  We adopted an a priori threshold of $<5\%$ for determining if two spectra were too dissimilar to use in a joint fit, and the resulting spectral groups can be seen in our preliminary spectral fitting results (Table~\ref{tab:tab2}).  We show typical $90\%$ confidence interval contours (fewest bins to contain $90\%$ of all samples) of two similar spectra from S24 in figure \ref{fig:MCMC}.  For a handful ($\sim3$) of instances when the behaviour of a source was not clear cut, i.e. spectrum $A$ and spectrum $B$ were different, while both were consistent ($>5\%$) with spectrum $C$ we took the more consistent of the two groupings, guided by the $90\%$ confidence contours (i.e. a noticeable difference in shape inside the parameter space between two groups).

\subsection{Spectral Fitting}
\label{s:res}
After identifying sets of spectra for each source, we proceeded to fit absorbed single-component models to the 0.5-8.0 keV spectra, following the prescription of \cite{2010ApJ...725.1805B}, who showed that the results from such fits -- particularly the behavior of absorption parameter $N_H$ -- can give a strong indication of the actual spectral state of a given source.  These results are summarised as a scheme for spectral fitting by means of a flow-chart in Figure 16 of \cite{2010ApJ...725.1805B}, and we direct the interested reader to inspect this chart in tandem with our results.  For thermally dominant spectra, where a disk blackbody component accounts for $>60\%$ of the source flux, show a level of absorption significantly above the Galactic value $N_H^{Gal}$ \citep{1990ARA&A..28..215D} when fit with an absorbed power law, while a power law dominated spectrum will have an absorption less than the Galactic value when fit with an absorbed disk blackbody.  If $N_H$ is significantly above the Galactic value for both fits, then the source was deemed to have intrinsic absorption. If $N_H$ was above $N_H^{Gal}$ for the power law fit but significantly below $N_H^{Gal}$ for the $diskbb$ model, then this indicates a thermal dominant spectrum with some power law component also present. Conversely, if $N_H$ from fitting was zero for both models, then this was indicative of the source spectrum being dominated by a steep ($\Gamma>1.7$) power law component with a cool ($kT_{in} < 0.5$ keV) disc component also present.   More specific effects are discussed in section~\ref{s:disc}.  

We present results of fitting $phabs \times powerlaw$ and $phabs \times diskbb$ in Table ~\ref{tab:tab2}, and indicate in the last column the spectral state(s) suggested by the \cite{2010ApJ...725.1805B} systematic scheme.  We emphasize that the results in this table demonstrate a phenomenological test.  Sources judged to be in a thermal dominant or power law dominant state were subsequently re-fit, imposing a lower limit on $N_H$ at $N_H^{Gal}$, and we refer the reader to Table~\ref{tab:res} for the realistic source properties.  The tendency for so many of the sources coincident with the dust lane to have an inferred 'intrinsic' absorption (i.e., $N_H$ from both absorbed power law and disk blackbody models was significantly above $N_H^{Gal}$), meant that we tried to determine the `true' absorping column along those lines-of-sight, for a more realistic and useful comparison.

\subsection{Dust lane sources}
There is clearly a strong correlation between sources that possess intrinsic absorption, according to our method, and coincidence with the dust lane, as expected if the sources lie behind or within the dust.  To define the spectral states for these sources, we obtain a value of $N_H$ independently from the X-ray spectral fitting, and consider the extra absorbing column in Cen~A as well as the Galactic contribution.  We converted a K-band optical depth map of the dust lane \citep{2009A&A...502L...5K} to $N_H$ assuming $A_K \sim 0.09A_v $ \citep{1992dge..book.....W} and $N_H\sim A_v\times 2.1\times 10^{21}~{\rm cm^{-2}}$ \citep{2009MNRAS.400.2050G}.  The mean $N_H$ was then calculated using the IRAF tool \emph{imexam}\footnote{http://stsdas.stsci.edu/cgi-bin/gethelp.cgi?imexamine} for each source coincident with the dust lanes, using circular apertures of radius $2\arcsec$, centered on the source position.  The $N_H^{DBB}$ and $N_H^{PO}$ from spectral fitting were compared to the mean $N_H$ inferred from the optical depth map at each source position, to infer the true spectral state, if possible.  Dust lane sources that we determined to be dominated by disk blackbody or power law spectra are included in Table~\ref{tab:res}, and in Figures \ref{fig:THERMAL} \& \ref{fig:POWER} when the parameters could be constrained.  A full discussion of these sources can be found in Section~ \ref{s:dustlane}.

\subsection{Inter-Observation Variability}
We can further our knowledge of these sources by taking the degree of inter-observational variability into account over the course of our 2 month snapshot.  We calculated the net photon flux\footnote{We emphasise that this does not involve assuming a spectral model} for each source region using the ciao tool \emph{aprates}\footnote{http://cxc.harvard.edu/ciao4.2/ahelp/aprates.html}, after calculating the average effective exposure in the region using a 0.5-8.0 keV exposure map.  To assess the variability, we calculate the fractional variability of the source, which we define for a set of fluxes $F_i$ as $(F_{max} - F_{min}) / F_{max}$.  For sources that are below the detection threshold in some observations, the $90\%$ upper-limit was used for $F_{min}$.  We present these resuls in Figure \ref{fig:vary}, where we show the fractional variability against spectral parameter ($kT_{in}$ \& $\Gamma$) for a given spectral state.  The majority of sources display $10-40\%$ variability over the time period spanned by the VLP.

\section{Discussion}
\label{s:disc}
In this section we first discuss each source coincident with the dust lane in turn, identifying the spectral state our method favors where possible.  We then present the unabsorbed luminosities as a function of key parameters of thermal and power law dominated states, and discuss the implications and source classification based on these results.  We discuss possible transient NS LMXBs in outburst and also those states for which we are not able to make an adequate diagnosis using simple models.  Finally, we offer an explanation for an apparent enhancement in the number of accreting BHCs found beyond the vicinity of the merged late-type galaxy.

\subsection{Dust lane sources}
\label{s:dustlane}
For those sources that are coincident with the dust lane and where spectral fitting suggests a high $N_H$ we infer the spectral state of a given source based on the mean absorption calculated from a K-band optical depth map.  Due to the extra level of consideration given to the dust lane sources,  we discuss each source in turn below, and where a particular state is suggested we present it in Table~\ref{tab:res}.  Our spectral diagnostic, inferring the state from fitting simple spectral models, is outlined in section \ref{s:res}.

\subsubsection{S4}
The brightest dust lane source, S4, is $\sim30\arcsec $ from the Cen~A nucleus, coincident with a southern filament of the dust.  From the optical depth map we calculate $N_H=(3.64\pm0.07) \times  10^{21}~{\rm cm^{-2}}$, which, while above the Galactic value, is still below $N_H^{DBB} (=6.5\times  10^{21}~{\rm cm^{-2}}$).  It seems likely that the spectrum is dominated by a thermal component; the discrepancy in absorption is either a consequence of additional intrinsic absorption from material local to the source, or the result of some significant non-thermal component contributing to the spectrum.  The true $kT_{in}$ may be somewhat cooler than that found from just fitting $phabs(diskbb)$, as suggested by the simulations of \cite{2010ApJ...725.1805B}, an effect we discuss in more detail for S12.  Fixing $N_H$ at the value obtained from the optical depth map preserves the cool disk temperature in spectral fitting.  There is a large fall in flux from this source, the spectral counts falling to 29 in the final observation from over 1100 in the first; however, it is still above our detection limit.  The low inner-disk temperature in the thermal state, combined with the high inter-observation variability, strongly suggests that this is a BH LMXB transient candidate, similar to S14 \citep{2012ApJ...749..112B}.  NS LMXBs at this luminosity, $L_x> 10^{38}~{\rm erg~s^{-1}}$, typically have a much flatter spectral shape over the \emph{Chandra} band.

\subsubsection{S7}
For S7 the absorption map gave $N_H=(6.52\pm 0.08) \times 10^{21}~{\rm cm^{-2}}$, which is below $N_H^{PO}$ but above $N_H^{DBB}$.  This suggests that the source is behind the dust lane in a thermal dominant state, but with some power law component present.  We are not able to determine the flux contribution of the power law component.  Based on the simulations of \cite{2010ApJ...725.1805B} we estimate the absolute systematic error $\pm 0.5$~keV on the temperature of the disk component. 

\subsubsection{S12}
In the case of S12, the inferred $N_H=(16.5 \pm 0.2)\times 10^{21}~{\rm cm^{-2}}$.  This is significantly above the values found from spectral fitting, and allows for the source to be located within the dust lane, rather than behind it.  If we assume that the source is behind the dust lane, i.e. has a line-of-sight absorption column of $N_H=(16.5 \pm0.2)\times 10^{21}~{\rm cm^{-2}}$, then we have fitting results that were not produced by the simulation work of \cite{2010ApJ...725.1805B}, where $N_H^{DBB}$ went to zero and $N_H^{PO}$ was less than the Galactic value, which was the result obtained when the simulated spectrum was a dominant, steep power law component with a cool disk also present.  Those simulations were based on a fairly low value of $N_H$, less than $1\times 10^{21}~{\rm cm^{-2}}$, and so the tendency for $N_H^{DBB}$ to go to zero for instances where $N_H^{PO} < N_H^{Galactic}$ is not surprising, as $N_H^{DBB}$ will nearly always  be less than $N_H^{PO}$ and so the size of the high-likelihood region of the parameter space is small.  


To investigate our ability to recover the parameters of a heavily absorbed cool disk plus steep power law spectrum, we carried out simulations in XSPEC.  We assumed a heavily absorbed ($N_H\sim1.6\times 10^{22}~{\rm cm^{-2}}$) cool disk ($kT_{in}\sim 0.4$ keV) and a steep  power law ($\Gamma\sim 2.6$).  Sets of 200 simulated spectra were produced, varying the $diskbb$ fraction of the total flux from 0.1 to 0.9.  We then fitted $phabs \times diskbb$ to these spectra to look at the effect of varying the initial spectral parameters on those recovered by fitting a simple, single component spectrum.  We found that the fits recovered the initial properties of the disk when the simulated spectrum was thermally dominated, but as the fraction of flux taken by the power law component increased in significance, so the recovered value of $N_H$ decreased, eventually reaching $N_H^{DBB}$ comparable with that found from fitting the spectra of S12.  A consequence of lower $N_H^{DBB}$ as the power law contribution increased was that the recovered inner-disk temperature rose to beyond 1.5 keV.  This fits neatly with the $phabs \times diskbb$ fit to S12, which has a high value of $kT_{in}$ (=${3.96}_{+0.87}^{-1.73}$~keV), implying an unphysically massive stellar BH. We take this result as an indication that our assumption about S12 is correct; that it is obscured fully behind the dust lane, with a line-of-sight $N_H>10^{22}~{\rm cm^{-2}}$, and has a spectrum dominated by a power law with some small disk contribution.  Further simulations suggest that high inner-disk temperatures ($>3$ keV) are also recovered with a less steep power law component ($\Gamma$<1.5).  This is clearly a complicated parameter space, the full properties of which are beyond the scope of this work to investigate.  It is worth noting that these results indicate a significant under-estimation in the flux of the source by just fitting simple models; the actual $L_x$ is probably $>6\times 10^{38}~ {\rm erg~s^{-1}}$, and could conceivably  $>10^{39}~ {\rm erg~s^{-1}}$, based on our simulations.  Fitting more complex (two component) models failed to give acceptable fits or successfully accommodate both components, with poorly constrained parameters.

S12 is analogous to S102 in NGC~3379 \citep{2010ApJ...725.1805B}, for which simulations indicated that the spectra were dominated by a steep power law with some contribution from a cool disk, while a successful fit was achieved using a combined $diskbb+powerlaw$ model yielding $kT_{in}\sim 0.14$ and $\Gamma\sim 1.6$ and $L_x=1.1 \times 10^{39} {\rm erg~s^{-1}}$.

\subsubsection{S13}
We calculate $N_H=(7.95\pm 0.13)\times 10^{21}~{\rm cm^{-2}}$ in the vicinity of S13, which is below or consistent with the $N_H^{PO}$ values obtained, but these are poorly constrained and the fits are statistically poor with $\chi_\nu^{2} > 1.3$.  Conversely, $N_H$ is above $N_H^{DBB}$, which is significantly larger than $N_H^{Gal}$.  As was the case with S7, such results are indicative of a thermally dominant state with some non-thermal emission also present, by which we mean that the thermal state contributes $> 60\%$ of the total flux.  The inner temperature in the disk of this source is less than that typically seen for NS systems at equivalent luminosities.  This source displays high inter-observation variability, but remains in outburst throughout the length of our observations.  This variability coupled with the low inner-disk temperature suggests that this system is a BHC LMXB, though examination of further observations will be required to show more substantive evidence of the transient nature of this source.

\subsubsection{S14}
S14 is a transient source, analysis of which has previously been reported in \cite{2012ApJ...749..112B}. $N_H^{PO}$ was larger than the value inferred from the dust lane, $N_H^{DBB}$ was found to be consistent with this value, which led us to conclude the source was in a thermally dominant state.  The cool disk at high luminosity, coupled with its transient nature, led us to conclude that the source is a BH LMXB candidate.  Our fitting results for the first group suggests the presence of a power law component in the more luminous state.

\subsubsection{S24}
While S24 is coincident with the dust lanes, its position is not covered by our optical depth map.  This being the case we cannot feasibly proceed further with our inference of spectral states based on the behavior of $N_H$, given that our first stage of spectral fitting is highly suggestive of increased line-of-sight absorption, above the Galactic value.

\subsubsection{S28}
For S28, we find $N_H=(7.63 \pm 0.12) \times 10^{21}~{\rm cm^{-2}}$ from the K-band optical depth map.  We fit two groups of spectra for this source.  For the first group, consisting of spectra from obsIDs 7797 and 8490, we find a large uncertainty on $N_H$ for both models, which is consistent with any location relative to the dust lanes at the $2\sigma$ level.  The second group, using spectra from obsIDs 7800 and 8489, have $N_H$ significantly above $N_H^{Galactic}$, which indicates that there is significant absorption along the line-of-sight to the source, which we attribute to the dust lane.  For this group, $N_H^{PO}$ is above the value obtained from the optical depth maps, while $N_H^{DBB}$ is consistent with this value, suggesting that the source is in a thermally dominant state.  Assuming that the source is behind the dust lane, as the second set of spectral fitting results suggests, and assuming a $1\sigma$ knowledge of $N_H^{PO}=0.51_{0.24}^{0.26}\times 10^{22}~{\rm cm^{-2}}$ and $N_H^{DBB}=0.44_{0.15}^{0.18}\times 10^{22}~{\rm cm^{-2}}$, then, for the first group, the situation is similar to that seen in S12, where $N_H^{PO}$ is consistent with the line-of-sight value while $N_H^{DBB}$ is lower but non-zero.  The higher, unrealistic (and poorly constrained), inner-disk temperature of 5.97 keV is consistent with this result, suggesting that the source is in a power law state during these epochs, with a minor contribution from a cool ($kT_{in}<0.5$ keV) disk component.

\subsubsection{S34}
S34 is close to the edge of the dust lane but is not coincident with the lane itself.  The high $N_H$ in one of the spectral fitting groups, of which obsID 8490 provided the only spectrum, prompted us to investigate the line-of-sight absorption using the optical depth map; we found $N_H=(1.10\pm0.08)\times 10^{21}~{\rm cm^{-2}}$, consistent with the Galactic value.  We are not able to deduce more about the state from this information; the source could have some intrinsic absorption not present in the other obsIDs.  This behavior is reminiscent of the black hole in globular cluster RZ 2109 from NGC~4472, where the variation in $N_H$ is believed to be the result of a photoionizing, high-velocity wind \citep{2010ApJ...721..323S}.  However, S34 is much fainter than this class of system, and the apparent increase in absorption could be an effect of incorrectly modelling the boundary layer emission, or the absorption may be a real effect and the system is at a high inclination, undergoing a long period of dipping.

\subsubsection{S38}
S38 is only detected in obsID 8490. From the optical depth map, we infer $N_H=(4.54\pm0.07)\times 10^{21}~{\rm cm^{-2}}$ for this source.  This is consistent with $N_H^{DBB}$ and significantly less than $N_H^{PO}$, suggesting a source in the thermally dominant state with an inner disk temperature $\sim0.65$ keV,  similar to S14, albeit at a lower luminosity of $L_x\sim5.5\times 10^{37}~{\rm erg~s^{-1}}$, which points to it being a transient BH LMXB candidate.

\subsubsection{S46}
The X-rays from S46 would only experience an absorption column of $N_H=(2.81 \pm 0.08) \times 10^{21}~{\rm cm^{-2}}$ if the source is behind the dust lane.  $N_H^{PO}$ is less than this line-of-sight value, and consistent with $N_H^{Galactic}$ while $N_H^{DBB}$ tends to zero, suggesting a source in front of the dust lane, in a power law dominated hard state.

\subsection{BH \& NS LMXBs}
None of the sources in our sample is confirmed to have $L_x > 4 \times 10^{38} ~{\rm erg~s^{-1}}$, compared to 3 sources apiece in NGC~3379 and NGC~4278 in the sample of \cite{2010ApJ...725.1824F}.   We caution that our sample is drawn from within the half-light radius, as opposed to $D_{25}$, on account of the relative proximity of Cen~A. In addition, Cen~A is a slightly smaller galaxy with $M_B=-21.1$ \citep{2007ApJ...654..186F} compared to $-22.28$ and $-22.02$ for NGC~3379 and NGC~4278 \citep{2006MNRAS.366.1126C}, and the LMXB population scales with galaxy mass to first order.  While we also exclude the sources that vary during the course of an observation or are close to the Cen~A nucleus for a later study, it is unlikely that these sources have $L_x > 5 \times 10^{38}~{\rm erg~s^{-1}}$ (Table \ref{tab:src}) unless the intra-observation variability is extreme.  Assuming the XLF follows an unbroken power law, with $dN/dL \sim K L^\alpha$ and $\alpha=-2$ \citep{2004ApJ...611..846K}, and assuming there are $\sim 40$ sources between $\sim2\times 10^{37}~{\rm erg~s^{-1}}$ and $\sim10^{38}~{\rm erg~s^{-1}}$ (Table~\ref{tab:src}), then $K\sim10^{39}$ and there would be $\sim 8$ sources between $10^{38}~{\rm erg~s^{-1}}$ and $5 \times 10^{38}~{\rm erg~s^{-1}}$ and we would expect a single source in  $5 \times 10^{38}~{\rm erg~s^{-1}}$ to $10^{39}~{\rm erg~s^{-1}}$. These values are consistent with our results.

\input{thermal_results}

\begin{figure*}[htb!]\center
{\includegraphics[width=1\hsize]{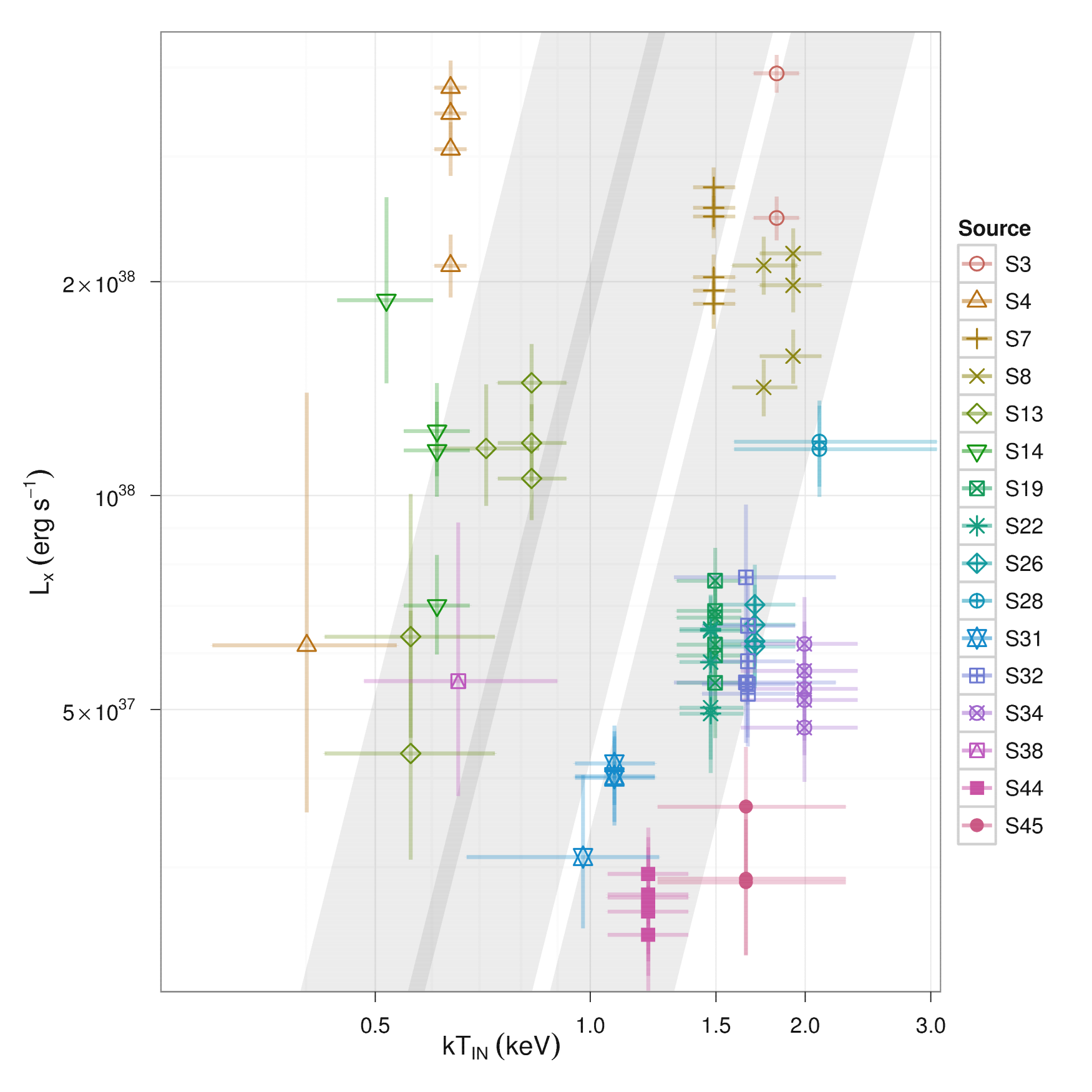}
\caption{X-ray luminosity against inner-disk temperature for states deemed to be in the thermal state.  Diagonal bands represent constant mass for (\emph{left} to \emph{right}) 10 ${\rm M_\odot}$, 5 ${\rm M_\odot}$ and 2 ${\rm M_\odot}$ non-rotating compact objects. The width of the bands shows an assumed inclination $\theta$ of $\cos\theta =0.25$ to $\cos\theta =0.75$. \label{fig:THERMAL}}}
\end{figure*}

\begin{figure}[htb!]\center
{\includegraphics[width=1\hsize, angle=270]{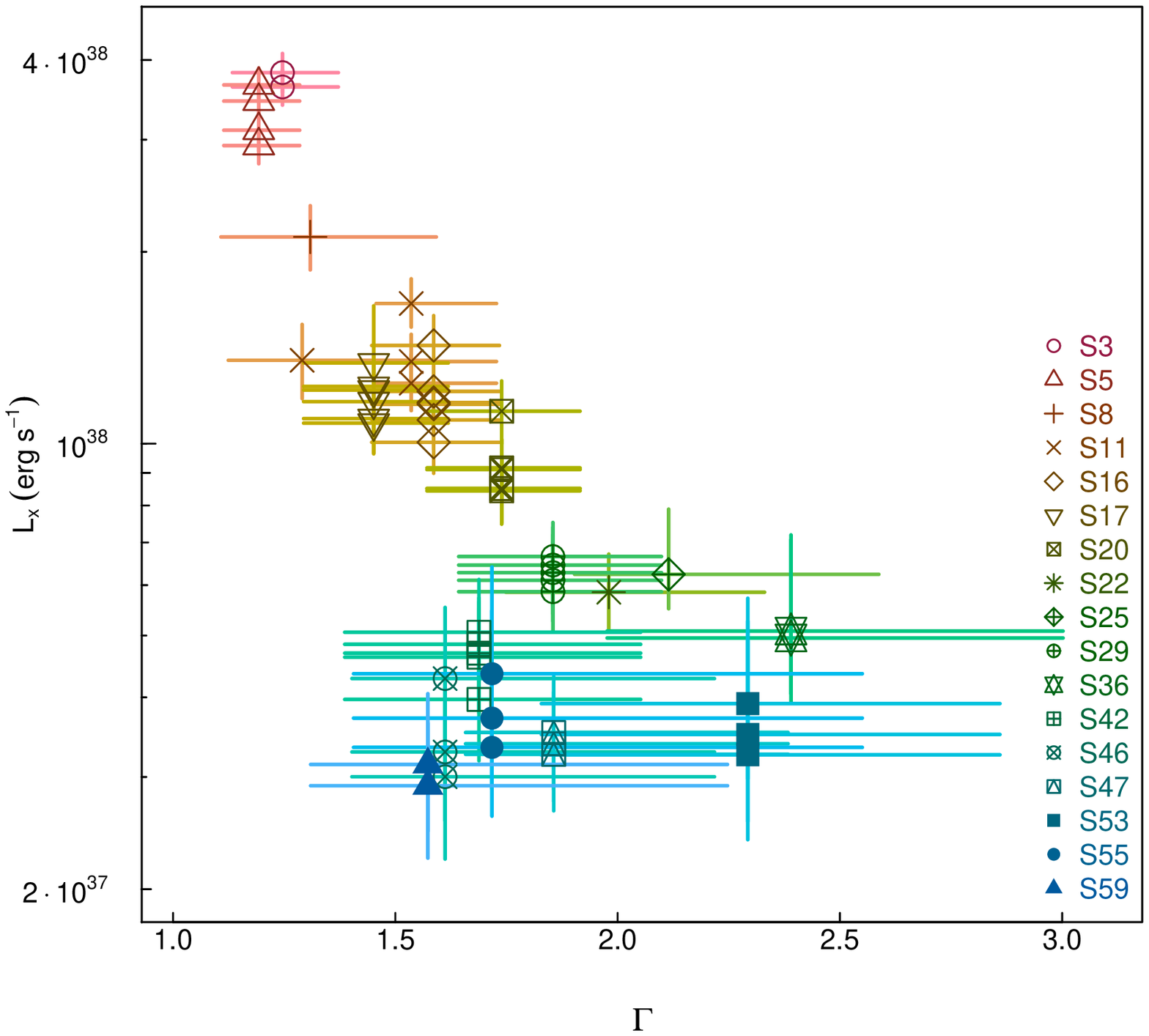}
\caption{X-ray luminosity against photon index for those states determined to be power law dominated. \label{fig:POWER}}}
\end{figure}

\begin{figure*}[htb!]\center
{\includegraphics[width=0.4\hsize]{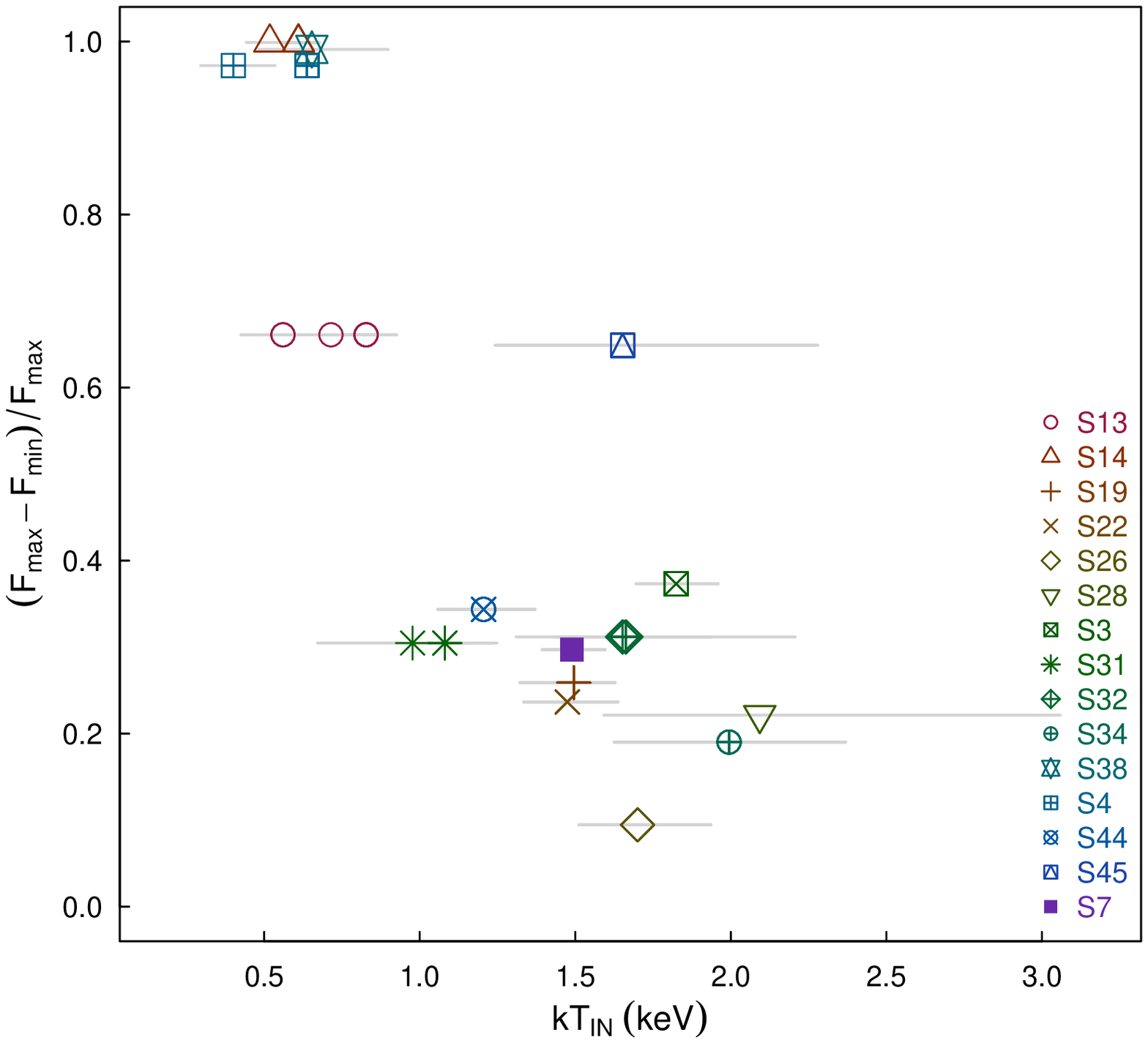} \includegraphics[width=0.4\hsize]{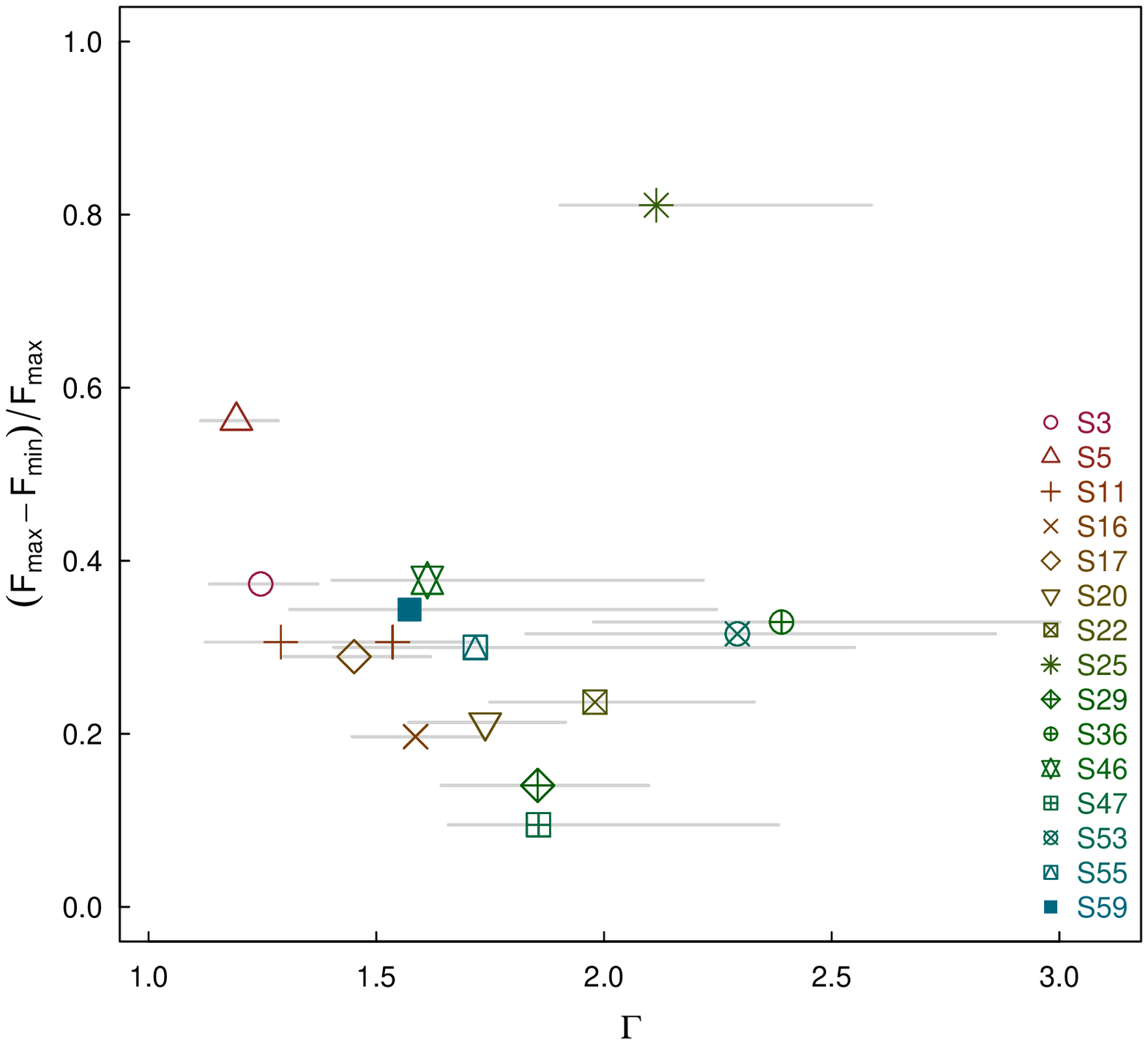}
\caption{Inter-observation fractional variability against \emph{(Left)} thermal state inner-disk temperatures and \emph{(Right)} Power law state photon indices. \label{fig:vary}}}
\end{figure*}

We present the thermally dominant and power law dominant spectral properties in Figures~\ref{fig:THERMAL} \& \ref{fig:POWER}, respectively.  In Figure ~\ref{fig:THERMAL}, we plot the unabsorbed disk luminosity against inner-disk temperature for the thermally dominant states (note that one point is representative of one ACIS spectrum, as we allowed for normalization, i.e. the flux, to be a free parameter during fitting), and we show illustrative bands of constant mass for $10M_\odot $, $5M_\odot$ and $2M_\odot$ assuming a non-rotating compact object \citep{2004MNRAS.347..885G}, the width of the bands shows the variation with inclination $\theta$, from $\cos\theta=0.25$ to $\cos\theta=0.75$ \citep[for which we use correction factors presented in ][]{1997ApJ...482L.155Z}, and we assume $f_{col}=1.8$.  What is most striking is the apparent bimodal nature of the thermal state sources, between those that are consistent with, or to the right of our $2M_\odot$ band, and those to the left of the $5M_\odot$ b.  In Figure 6 of \cite{2012ApJ...749..112B}, we presented a population of Local Group thermal states, all of which lie to the left of our supposed $2M_\odot$ line.  The fact that the bulk of our sources lie to the right of this line, with $kT_{in} > 1 $ keV in the range of $L_x\sim 10^{37}-10^{38} ~{\rm erg~s^{-1}}$, is very suggestive that our thermal state sample is dominated by NS LMXBs.   

We suggest tentatively that this result is reminiscent of the well-known `mass gap' problem \citep{1998ApJ...499..367B}.  The mass distribution of compact objects in transient systems strongly deviates from theoretical predictions, with a characteristic paucity between the most massive NSs ($\sim3 M_\odot$) and the least massive BHs, while the mass distribution of BHs is seen to peak at $\sim8 M_\odot$. If not a systematic effect \citep[e.g.][]{2011ApJ...741..103F,2012ApJ...757...36K}, then this feature in the mass distribution favours `rapid' initial stellar collapse models of supernovae, where initial instabilities grow on timescales of 10-20 ms \citep{2012ApJ...757...91B}, over models that require more prolonged instability growth ($>200$ ms).  To quantitatively test if these data are distributed bimodally in this plane, we first calculate a `mass' for each source based on the observed peak luminosity, assuming $\cos\theta=0.5$.  We then fitted\footnote{Using the function \emph{fitdistr}, part of the MASS package in R} both a single Gaussian, and then dual Gaussians to the observed mass distribution and from these fits calculate the Akaike's Information Criterion (AIC) \citep{1974ITAC...19..716A} for both models.  The dual Gaussian fit had means (standard deviations) of 1.93 (0.76) $M_\odot$ and 15.74 (8.46) $M_\odot$ while the single Gaussian was centred on 5.6 (7.55) $M_\odot$.  The AIC for the single Gaussian fit (=114.11) was larger than that of the dual Gaussian fit (=83.66), such that the latter was strongly favoured, with the single Gaussian $ 2.44\times 10^{-7}$ times as probable as the double Gaussian.    These results are the first hint that the mass gap exists outside the Local Group.

BH systems do not have thermal emission from a boundary layer, as one would expect from the NS LMXBs.  The flatness of the S3 and S5 spectra, which our method suggests are power law dominated with $\Gamma~\sim 1.2$, is more consistent with the spectra of LMXB NS Z-track sources than those of BH sources in the hard state at $L_x > 10^{38}~{\rm erg~s^{-1}}$.  We know this based on simulating the spectra of NS LMXBs using the model and parameters reported by \cite{2012arXiv1207.1107L}, who modeled the spectra of a bright NS LMXB, GX17+2, using a combination of disk blackbody, blackbody, and a power law component that is significant only on the Z-track horizontal branch.  Regardless of whether this is a physically correct description or not,  it is clearly a good phenomenological description of the spectral shape.  At similar count rates to S3 and S5, we find that the parameters of this multi-component model are poorly recovered by a  two-component $phabs(diskbb+bbody)$ fit, but that the shape of the spectrum is typically well-described by a shallow power law of $\Gamma \sim 1.0-1.4$, which does not vary substantially between Z-track states over the \emph{Chandra} bandpass.  For what is judged to be the thermally dominant state of S3, $kT_{in}$ is higher than for the BH candidate systems.  We therefore conclude that S3 and S5 are candidate Z-track sources.   

The power law states do not show a bimodality as we see for the thermal states, but they are consistent with Galactic NS and BH XBs, and there appears some slight favoring of steeper $\Gamma$ at lower luminosities.  However, such an effect  was also seen by \cite{2010ApJ...725.1824F}, whose limiting luminosity for spectral fitting was much higher.   In Figure 9 of \cite{2010ApJ...725.1805B} it is shown using 1000 count simulated spectra that the change in $\Gamma$ with flux contribution favoured  larger values from spectral fitting, particularly for cooler disks.  It is conceivable that this effect becomes more pronounced when there are fewer counts. 
 
\subsubsection{NS Transient Candidates}
Of the less luminous sources that do not appear to be BHCs we find that S25, S35 and S45 display extreme variability. S25 is only bright enough for spectral fitting in one observation, where we determine the source to be in a power law dominant state.  While the source is always above our detection limit, the flux falls to $\sim 20 \%$ of the initial value after obsID 7797, and the spectrum becomes dramatically softer by obsID 8490, with $log_{10}(S/H) \sim 1.1$ and a $90\%$ confidence lower-limit of $\sim0.6$ in obsID 8490.  That a source displays quasi-super-soft behavior \citep{2003ApJ...592..884D} at low luminosity but $\Gamma \sim 2$ may indicate that S25 is a classical nova.  

S35 is only detected in obsID 8490, but  in fitting simple models we are not able to constrain $N_H$ such as to show that it is above, below or consistent with $N_H^{Gal}$.  All the parameters are poorly constrained, but taken at face value the $\Gamma$ or $kT_{in}$ are not inconsistent with NS LXMBs.  The variability demonstrated by the source would be consistent with a transient atoll-type NS LMXB towards the peak of outburst.  

We believe that S45 may also be a transient NS LMXB in outburst.  It displays significant temporal variability over the course of the six observations, and appears to be in a thermally dominant state with an inner-disc temperature $kT_{in} \sim 1.3-2.3$~keV, consistent with disc temperatures obtained from Aql X-1 \citep[e.g. ][]{2007ApJ...667.1073L}, which we suggest is an analogue to this source.

\subsection{S48: A high magnetic field NS in a GC?}
\label{sec:S48}
S48 is a peculiar source. Coincidence with a spectroscopically-confirmed GC means that it is highly unlikely to be a background AGN or foreground star.  The spectrum is well-fit by a power law model of $\Gamma\sim 0.7 $ experiencing negligible absorption, and a good fit cannot be achieved for the thermal model (Table~\ref{tab:res}). The luminosity of S48 makes it unlikely to be a cataclysmic variable, but it is similar to IGR J17361-4441, a hard X-ray transient detected by INTEGRAL in the Galactic globular cluster NGC~6388 \citep{2011ATel.3566....1F}, which is possibly a high-magnetic field binary such as GX 1+4 \citep{2005ApJ...627..915P}, and has a similar power law slope.  The spectra of these sources are well described by a cut-off power law of $\Gamma\sim 0.7-1.0$ with a high energy cut-off at 25 keV \citep{2011A&A...535L...1B}. The spectral shape is consistent with the compact object being a highly magnetized NS, the emission emanating from accretion columns that impact the NS surface.  Young, highly magnetized radio pulsars have been observed in GCs by \cite{2011ApJ...742...51B}, so this is a plausible explanation.  However, such sources typically show short-term modulation on a scale of a few hours \citep{2008ApJ...675.1424C}; while we do not detect any variability inside of any observation (Table  \ref{tab:src}), such modulation might be too small to be detected.

\subsection{Uncharacterised Spectral States}
\label{sec:uncha}
There are a handful of examples of spectra that could not be formally distinguished by our method (Section \ref{s:res}).  In the case of S29, S32, S34 and S36, we are able to determine the spectral state for a subset of the observations but not for others that we had previously shown to have a different spectral shape.  In the case of S29, obsID 7798, the source may be experiencing an increase in absorption or some intermediate stage as it moves away from the hard state, as is the case with S34 obsID 8490.  In obsIDs 7797, 7798 and 8490 of S36, we find $N_H$ consistent with the $N_H^{Gal}$ for both models.  With the point estimate of $N_H^{DBB}$ much closer to zero, we could tentatively suggest that this source remains in a power law dominated state throughout, the spectra becoming harder at certain times.  We note that the softer state of this source coincides with its most luminous epoch, reminiscent of Galactic NS LMXBs such as 4U 1635-536, the spectra of which tend to harden as the source reaches its lowest outburst luminosities.  In S31, there are two clear examples that seem to favor a thermal shape, but there is also clear evidence of a much harder state in 0bsIDs 7798 \& 8489 where it becomes more difficult for the fit to constrain the absorption column. For S32 obsID 8490, $N_H$ could not be constrained for either simple model.  

\subsection{A different population?}
Spectral fitting of simple models indicates that 6 of the 8 sources we highlight as being coincident with the dust lane (Table \ref{tab:src}) experience a greater line-of-sight absorption than that provided by Galactic column alone.  The two sources that were consistent with $N_H^{Gal}$, S34 and S46, are on the edge of the dust lane, they are thus the worst candidates for coincidence and were labeled as such out of caution.  Of sources not coincident with the dust lane, only one source, S39, appears to experience excess absorption by our diagnostic.  Inspection of the spectra of this source in more detail has led us to conclude that it is a background AGN, with a redshifted Fe line at $\sim 4$ keV present in the spectrum.

The good agreement between sources coincident with the dust lane and those that possess absorption surplus to the Galactic column strongly suggests that these sources experience additional absorption from the dust lane, and validates our previous analysis in section~\ref{s:disc} using a K-band optical depth map to estimate of the average $N_H$ in the vicinity of each source (with the exception of S24, which is not within the FoV of the map), and then infer the dominant component in each of the spectra.  This technique assumes that the sources are completely behind the dust lane, and that the $N_H$ does not vary substantially over a $2\arcsec$  radius region, while in reality the material is likely to be clumpy and non-uniformly distributed.

Our four transient BH LMXB candidates are all dust lane sources, and are observed with $kT_{in} < 0.9$ keV. Two further dust lane systems, S12 \& S28, possess very cool ($<0.4$ keV) disks with a strong power law component, mildly suggesting that they also possess BHs. This result indicates an enhancement of classic BHC transients inside of the merged late-type galaxy relative to the halo.We note that previous studies identified the two known ULXs in the south-west quadrant of Cen~A as transient X-ray binaries that are strong BHCs \citep[e.g.,][]{2006ApJ...640..459G,2008ApJ...677L..27S}. Although their presence strongly suggest that there is a population of transient BHCs outside of the merger region, neither source would have been identified as a transient based on the observations we consider in this paper. As such, we exclude them from our analysis.

Previously, hardness-intensity or color-color diagrams have been used to indicate which sources in a sample are heavily absorbed or not, but these were not ideal for giving an indication as to whether the sources themselves were similar, as any further difference in hardness or color could be ascribed to the additional (unknown and unconstrained) absorption.  These results from detailed spectral fitting clearly point to particular types of source being present in the late-type galaxy that are absent from the rest of Cen~A.  

We propose a straightforward explanation of the lack of BHC transients outside of the late-type remnant.  An empirical relation  exists between orbital period and peak outburst luminosity \citep{2010ApJ...718..620W}, with larger luminosities produced by the systems with longer orbital periods.  For transients that contain BHs, this luminosity exceeds $10^{38}~{\rm erg~s^{-1}}$, which  corresponds to an orbital period of $\sim10$ hours, based on the \cite{2010ApJ...718..620W} relation.  For a main sequence companion filling its Roche lobe, the mass $m$ in ${\rm M_\odot}$ will be related to the orbital period $P$ in hours by $m\sim0.11P$ \citep{2002apa..book.....F}, which means that the companion star must be be at \emph{least} 1 ${\rm M_\odot}$.  It is likely that the donor is more evolved, particularly in a BH LMXB transient \citep[$<0.75 M_{MS}$, ][]{1996ApJ...464L.127K}, and therefore the actual main sequence lifetime of the $\sim1 M_\odot$ donors is shorter.  The majority (70-80\%) of stars in the halo of Cen~A were formed $12\pm1$ Gyr ago, with the rest formed from some event 8-10 Gyr later \citep{2011A&A...526A.123R}.  Therefore the majority of stars in the halo have $m < 1 {\rm M_\odot}$, which accounts for the relative absence of bright BH LMXB transients.  Inside the late-type galaxy the star formation rate is at least 0.1 ${\rm M_\odot~yr^{-1}}$, though could be as high as  1 ${\rm M_\odot~yr^{-1}}$ \citep{2000ApJ...528..276M}; therefore there is a population of $\geq 1 {\rm M_\odot}$ donors to feed the accreting BHs by Roche lobe overflow.  


It has been proposed that the two ULXs in Cen~A are BH LMXBs observed in the so-called \emph{ultraluminous} state \citep[e.g.,][]{2009MNRAS.397.1836G} at $L_x> 10^{39} {\rm erg~s^{-1}}$ and in quiescence with $L_x<10^{36}~{\rm erg~s^{-1}}$. These sources have not been observed in the `classic' BH LMXB states \citep{2006ARA&A..44...49R} with $L_x ~ 10^{37}$ -- $10^{39}~{\rm erg~s^{-1}}$; the non-UL outburst luminosity of these sources is unknown  and therefore they should not be compared directly with the dust lane BHCs until they have been observed below $10^{39}~{\rm erg~s^{-1}}$ but above the detectability threshold.  In these relatively atypical sources, it is not yet clear how the orbital period and peak outburst luminosity are related, which complicates direct comparison with the dust lane sources.

If confirmed, the above explanation for the lack of BHC transients outside of the late-type remnant also accounts for the results of \cite{2010ApJ...721.1523K} for a sample of other nearby early-type galaxies, who found the population of sources with $L_x > 4 \times 10^{38} {\rm erg~s^{-1}}$ diminished significantly with early-type age, a steepening of the discrete source XLF.  We recommend population synthesis work be carried out to investigate the effects of a declining population of transient BH LMXBs with stellar age.  We also suggest a deep monitoring campaign of other early-type galaxies with Chandra, to further investigate occurances of transient BHC XBs.

\section{Conclusions}
Our investigation into the X-ray binaries of Cen~A has found:

i) The population is mostly NS LMXBs, with 6 BH candidates, four of which are inferred as such from the inner-disk temperature in the thermal state and inter-observational variability, and two from the inferred presence of a cool disk at a high luminosity.  The proximity of Cen~A, coupled with the depth of the VLP data, means that this is the only early-type galaxy where it is possible to perform detailed spectral fitting to sources in the classic XB luminosity range of $10^{37}-10^{38}~{\rm erg~s^{-1}}$, similar to those sources we observe in the Milky Way.  As a population of NS LMXBs outside of the Local Group, these are older analogues of sources that have been studied since the dawn of X-ray astronomy.  We identify two sources that we suggest are Z track systems, based on their spectral shape, unabsorbed luminosity and persistent nature. 

ii) There is some evidence that the mass distribution of compact objects is bimodal, and there is tentative evidence of the so-called `mass gap' between $\sim 2.5-5 M_\odot$ that has been observed in the population of Galactic transient LMXBs.  This is the first time that the mass gap has been hinted at outside the Local Group.  

iii) Besides the two known ULXs in Cen~A, the only black hole candidate transient systems we can identify are found coincident with the dust lanes that arise from the merged late-type galaxy.  These are four transient BH LMXB candidates observed in the thermal-dominant state.  We propose that this is simply the result of stellar age; the older population of stars in the halo is unable to supply Roche lobe-filling companions that are required for the transient BHCs to have outburst peak $L_x > 10^{38}{ \rm erg~s^{-1}}$. This also explains other recent results that show a steepening of the XLF of early-type galaxies with stellar age for $L_x > 4-5 \times 10^{38}{ \rm erg~s^{-1}}$\citep{2010ApJ...721.1523K}, as such sources would require more massive companions that have long since evolved off the main sequence.  Similar analysis applied to other early-type galaxies where the ages of the stellar populations are known, coupled with stellar population synthesis work, is required to investigate the robustness of this explanation. 

\acknowledgments
This work was supported by NASA grant NAS8-03060. MJB thanks SAO and the University of Birmingham for financial support.  RV is supported by NWO Vidi grant 016.093.305.  GRS acknowledges the support of an NSERC Discovery Grant. We also thank Rudy Wijnands, Vinay Kashyap, Jeff McClintock, Andrea Prestwich, Rob Barnard, Kim Dong-Woo and Alastair Sanderson for useful discussions.  Finally, we thank the anonymous referee for their careful consideration of our work.


\bibliography{popbib}{}
\bibliographystyle{apj}


\clearpage
 \LongTables
\input{tabtest}

\end{document}

%% file: sourcedata.tex
\pagebreak
\clearpage
\begin{deluxetable}{lccccccccccr}
\tabletypesize{\footnotesize}
\tablewidth{0pt} 
\tablecaption{Basic Source Properties \label{tab:src}}
\tablehead{
&  &  & \multicolumn{6}{c}{Net Counts (0.5 -- 8.0 keV)} & & &  \\
\cline{4-9} \\
\colhead{Source} &
\colhead{$\alpha$} & 
\colhead{$\delta$}&  
\colhead{7797} & 
\colhead{7798} & 
\colhead{7799} &  
\colhead{7800} &   
\colhead{8489} & 
\colhead{8490} & 
\colhead{${\rm G-L_{max}}$} &
\colhead{Type} &
\colhead{Notes} \\ 
} 
\tablecolumns{11} 
\startdata
S1      &$ 13^{h} 25^{m} 18.25^{s} $&$ -43^{\circ} 03\arcmin 04\farcs9 $ &  $7183_{139}^{279}$  & $8366_{151}^{302}$  & $7372_{142}^{284}$  & $6623_{134}^{270}$  & $9624_{161}^{323}$  & $8454_{151}^{303}$ & 5 & & \cite{2008ApJ...677L..27S} \\ 
S2      &$ 13^{h} 25^{m} 38.29^{s} $&$ -43^{\circ} 02\arcmin 05\farcs6 $ &  $1685_{68}^{136}$  & $1364_{61}^{122}$  & $1429_{62}^{125}$  & $1062_{54}^{108}$  & $1858_{71}^{142}$  & $1366_{61}^{122}$ & 9 & &\\ 
S3      &$ 13^{h} 25^{m} 02.70^{s} $&$ -43^{\circ} 02\arcmin 43\farcs3 $ &  $1120_{55}^{110}$  & FoV  & FoV  & $917_{51}^{103}$  & $1575_{66}^{132}$  & $981_{52}^{103}$ & 1 & NSC \\ 
S4      &$ 13^{h} 25^{m} 26.16^{s} $&$ -43^{\circ} 01\arcmin 32\farcs6 $ &  $1101_{57}^{114}$  & $929_{53}^{107}$  & $897_{52}^{105}$  & $555_{43}^{87}$  & $197_{28}^{56}$  & $29_{17}^{35}$ & 0 & TBHC & D \\ 
S5      &$ 13^{h} 25^{m} 07.64^{s} $&$ -43^{\circ} 01\arcmin 15\farcs3 $ &  $880_{49}^{98}$  & $174_{23}^{46}$~C  & $176_{23}^{46}$~C  & $726_{46}^{93}$  & $1009_{53}^{105}$  & $1040_{53}^{106}$ & 5 & NSC & GC0129 \\ 
S6      &$ 13^{h} 25^{m} 06.30^{s} $&$ -43^{\circ} 02\arcmin 21\farcs1 $ &  $741_{45}^{90}$  & FoV & FoV  & $614_{43}^{86}$  & $1024_{53}^{106}$  & $613_{41}^{82}$ & 6 & & \\ 
S7      &$ 13^{h} 25^{m} 23.69^{s} $&$ -43^{\circ} 00\arcmin 09\farcs5 $ &  $756_{46}^{91}$  & $787_{48}^{96}$  & $871_{50}^{101}$  & $655_{44}^{87}$  & $608_{41}^{82}$  & $603_{41}^{82}$ & 0 & NSC & D\\ 
S8      &$ 13^{h} 25^{m} 22.88^{s} $&$ -43^{\circ} 01\arcmin 24\farcs9 $ &  $818_{48}^{96}$  & $703_{47}^{94}$  & $557_{41}^{83}$  & $829_{50}^{100}$  & $618_{42}^{85}$  & $651_{43}^{86}$ & 0  & NSC &  \\ 
S9      &$ 13^{h} 25^{m} 28.19^{s} $&$ -43^{\circ} 00\arcmin 56\farcs5 $ &  $539_{48}^{97}$  & $684_{51}^{102}$  & $701_{51}^{103}$  & $747_{52}^{104}$  & $431_{45}^{91}$  & $368_{47}^{95}$ & 0 & & N \\ 
S10     &$ 13^{h} 25^{m} 27.45^{s} $&$ -43^{\circ} 02\arcmin 14\farcs1 $ &  $600_{41}^{83}$  & SC  & SC  & SC  & $537_{39}^{78}$  & $451_{36}^{72}$ & 8 &  &  \\ 
S11     &$ 13^{h} 25^{m} 54.57^{s} $&$ -42^{\circ} 59\arcmin 25\farcs4 $ &  $408_{35}^{69}$  & $278_{28}^{55}$~C  & $313_{29}^{58}$~C  & $353_{31}^{62}$  & $402_{33}^{67}$  & $533_{40}^{80}$ & 0 & NSC & GC0330 \\ 
S12     &$ 13^{h} 25^{m} 32.45^{s} $&$ -43^{\circ} 01\arcmin 34\farcs2 $ &  $483_{37}^{75}$  & $496_{39}^{78}$  & $440_{37}^{74}$  & $289_{31}^{62}$  & $448_{36}^{72}$  & $291_{30}^{59}$ & 1 & PBHC & D \\ 
S13     &$ 13^{h} 25^{m} 25.76^{s} $&$ -43^{\circ} 00\arcmin 55\farcs8 $ &  $451_{38}^{76}$  & $445_{42}^{85}$  & $419_{41}^{83}$  & $208_{34}^{68}$  & $159_{27}^{53}$  & $418_{37}^{75}$ & 0 & TBHC & D \\ 
S14     &$ 13^{h} 25^{m} 27.58^{s} $&$ -43^{\circ} 00\arcmin 23\farcs3 $ &  $407_{34}^{69}$  & $409_{37}^{74}$  & $421_{37}^{75}$  & $292_{33}^{66}$  & $19_{12}^{25}$  & $1_{1}^{14}$ & 0 & TBHC & D, \cite{2012ApJ...749..112B} \\ 
S15     &$ 13^{h} 25^{m} 26.43^{s} $&$ -43^{\circ} 00\arcmin 54\farcs2 $ &  $309_{34}^{68}$  & $271_{33}^{66}$  & $322_{35}^{70}$  & $225_{32}^{64}$  & $147_{28}^{55}$~R  & $417_{39}^{78}$ & 0 &  & N \\ 
S16     &$ 13^{h} 25^{m} 31.59^{s} $&$ -43^{\circ} 00\arcmin 03\farcs0 $ &  $303_{29}^{59}$  & $327_{31}^{63}$  & $338_{32}^{64}$  & $400_{34}^{68}$  & $353_{31}^{63}$  & $409_{35}^{70}$ & 1 & NSC & GC0225 \\ 
S17     &$ 13^{h} 25^{m} 12.89^{s} $&$ -43^{\circ} 01\arcmin 14\farcs6 $ &  $405_{33}^{67}$  & $311_{31}^{63}$  & $288_{30}^{61}$  & $323_{32}^{64}$  & $349_{31}^{63}$  & $84_{15}^{31}$ & 0 & NSC  & \\ 
S18     &$ 13^{h} 25^{m} 27.50^{s} $&$ -43^{\circ} 01\arcmin 28\farcs1 $ &  $368_{38}^{76}$  & $323_{40}^{81}$  & $370_{42}^{83}$  & $345_{41}^{83}$  & $355_{37}^{74}$  & $364_{37}^{75}$ & 0 & & N \\ 
S19     &$ 13^{h} 25^{m} 27.08^{s} $&$ -43^{\circ} 01\arcmin 59\farcs2 $ &  $335_{32}^{64}$  & $274_{35}^{71}$  & $317_{37}^{75}$  & $194_{32}^{64}$  & $363_{33}^{66}$  & $280_{29}^{59}$ & 0 & NSC  & \\ 
S20     &$ 13^{h} 25^{m} 28.75^{s} $&$ -42^{\circ} 59\arcmin 48\farcs3 $ &  $287_{29}^{58}$  & $301_{31}^{62}$  & $341_{32}^{65}$  & $332_{31}^{63}$  & $329_{30}^{61}$  & $256_{28}^{56}$ & 0 & NSC & \\ 
S21     &$ 13^{h} 25^{m} 22.32^{s} $&$ -42^{\circ} 57\arcmin 17\farcs3 $ &  $282_{28}^{57}$  & $172_{24}^{48}$  & $330_{31}^{63}$  & $244_{26}^{53}$  & $286_{29}^{57}$  & $312_{31}^{62}$ & 10  &  & \\ 
S22     &$ 13^{h} 25^{m} 40.54^{s} $&$ -43^{\circ} 01\arcmin 14\farcs9 $ &  $265_{28}^{55}$  & $258_{27}^{54}$  & $312_{29}^{59}$  & $217_{25}^{50}$  & $112_{18}^{35}$  & $307_{30}^{59}$ & 1 & NSC  & \\ 
S23     &$ 13^{h} 25^{m} 35.50^{s} $&$ -42^{\circ} 59\arcmin 35\farcs2 $ &  $186_{23}^{47}$  & $271_{28}^{55}$  & $280_{28}^{56}$  & $308_{29}^{59}$  & $295_{29}^{57}$  & $256_{28}^{56}$ & 6 & &  GC0249 \\ 
S24     &$ 13^{h} 25^{m} 11.99^{s} $&$ -43^{\circ} 00\arcmin 10\farcs7 $ &  $288_{28}^{56}$  & $270_{30}^{61}$  & $256_{30}^{60}$  & $256_{29}^{58}$  & $259_{27}^{54}$  & $129_{19}^{38}$ & 1 & & D \\ 
S25     &$ 13^{h} 25^{m} 09.56^{s} $&$ -43^{\circ} 05\arcmin 29\farcs2 $ &  $286_{29}^{57}$  & FoV  & FoV  & SC  & $65_{16}^{32}$  & $53_{13}^{26}$ & 1 & NSC &  \\ 
S26     &$ 13^{h} 25^{m} 46.58^{s} $&$ -42^{\circ} 57\arcmin 03\farcs1 $ &  $264_{28}^{57}$  & $247_{26}^{52}$  & $284_{28}^{56}$  & $272_{27}^{55}$  & $92_{17}^{34}$  & $226_{26}^{53}$ & 2 & NSC & GC0295\\ 
S27     &$ 13^{h} 25^{m} 43.97^{s} $&$ -43^{\circ} 06\arcmin 06\farcs8 $ &  $273_{28}^{56}$  & $211_{25}^{51}$  & $248_{27}^{55}$  & $70_{19}^{37}$  & $202_{24}^{48}$  & $206_{24}^{49}$ & 6 &  &  \\ 
S28     &$ 13^{h} 25^{m} 24.19^{s} $&$ -42^{\circ} 59\arcmin 59\farcs4 $ &  $269_{28}^{56}$  & SC  & SC  & $226_{26}^{52}$  & $271_{28}^{56}$  & $213_{25}^{51}$ & 0 & PBHC & D \\ 
S29     &$ 13^{h} 25^{m} 33.93^{s} $&$ -42^{\circ} 58\arcmin 59\farcs7 $ &  $217_{25}^{50}$  & $217_{25}^{50}$  & $232_{26}^{52}$  & $215_{24}^{49}$  & $257_{27}^{54}$  & $224_{26}^{52}$  & 0 & NSC  &  \\ 
S30     &$ 13^{h} 25^{m} 12.03^{s} $&$ -43^{\circ} 00\arcmin 44\farcs7 $ &  $254_{26}^{53}$  & SC & SC  & R  & $228_{25}^{51}$  & $81_{15}^{30}$ & 6  &  & \\ 
S31     &$ 13^{h} 25^{m} 23.57^{s} $&$ -43^{\circ} 02\arcmin 20\farcs5 $ &  $246_{26}^{53}$  & $209_{29}^{58}$  & $161_{27}^{54}$  & $219_{29}^{58}$  & $215_{25}^{50}$  & $242_{26}^{52}$ &0 &NSC  & \\ 
S32     &$ 13^{h} 25^{m} 09.18^{s} $&$ -42^{\circ} 58\arcmin 59\farcs3 $ &  $249_{26}^{53}$  & $214_{26}^{53}$  & $177_{24}^{48}$  & $209_{26}^{52}$  & $207_{25}^{50}$  & $123_{19}^{38}$ & 0 & NSC & GC0134 \\ 
S33     &$ 13^{h} 25^{m} 23.52^{s} $&$ -43^{\circ} 01\arcmin 38\farcs4 $ &  $227_{26}^{52}$  & SC  & SC  & SC  & $236_{26}^{52}$  & $243_{26}^{53}$ &  7 &   & \\ 
S34     &$ 13^{h} 25^{m} 33.34^{s} $&$ -43^{\circ} 00\arcmin 52\farcs7 $ &  $208_{26}^{51}$  & $208_{27}^{54}$  & $192_{26}^{52}$  & $193_{26}^{52}$  & $224_{26}^{51}$  & $219_{27}^{53}$ & 0 & NSC & D \\ 
S35     &$ 13^{h} 25^{m} 18.50^{s} $&$ -43^{\circ} 01\arcmin 16\farcs1 $ &  $4_{4}^{9}$  & SC  & SC & SC  & $9_{8}^{18}$  & $218_{25}^{50}$ & 1 & T- & GC0182 \\ 
S36     &$ 13^{h} 25^{m} 32.01^{s} $&$ -43^{\circ} 02\arcmin 31\farcs3 $ &  $217_{25}^{50}$  & $207_{26}^{52}$  & $212_{26}^{52}$  & $188_{25}^{51}$  & $84_{16}^{31}$  & $153_{21}^{42}$ & 2 & NSC  & \\ 
S37     &$ 13^{h} 25^{m} 29.45^{s} $&$ -43^{\circ} 01\arcmin 08\farcs1 $ &  $162_{30}^{60}$  & $90_{37}^{74}$  & $146_{38}^{77}$  & $19_{19}^{50}$  & $213_{30}^{60}$  & $153_{30}^{60}$ & 0 &  & N \\ 
S38     &$ 13^{h} 25^{m} 22.85^{s} $&$ -43^{\circ} 00\arcmin 17\farcs4 $ &  $2_{2}^{10}$  & $9_{9}^{24}$  & $3_{3}^{19}$  & $0_{0}^{15}$  & $1_{1}^{9}$~R & $194_{24}^{48}$ & 0 & BHC & D \\ 
S39     &$ 13^{h} 25^{m} 28.42^{s} $&$ -43^{\circ} 03\arcmin 15\farcs2 $ &  $162_{21}^{43}$  & $187_{25}^{50}$  & $163_{24}^{47}$  & $171_{25}^{49}$  & $58_{13}^{26}$  & $110_{18}^{35}$ & 1 & AGN \\ 
S40     &$ 13^{h} 25^{m} 26.95^{s} $&$ -43^{\circ} 01\arcmin 04\farcs8 $ &  $171_{33}^{66}$  & SC  & SC  & SC  & $13_{13}^{38}$~R  & $186_{32}^{65}$ & 0 &   & N  \\ 
S41     &$ 13^{h} 25^{m} 26.94^{s} $&$ -43^{\circ} 00\arcmin 52\farcs5 $ &  $136_{28}^{56}$  & $162_{29}^{57}$  & $135_{27}^{55}$  & $164_{30}^{60}$  & $120_{27}^{54}$  & $182_{31}^{62}$ & 1 & &  N \\ 
S42     &$ 13^{h} 25^{m} 16.40^{s} $&$ -43^{\circ} 02\arcmin 55\farcs1 $ &  $190_{23}^{46}$  & $133_{24}^{48}$  & $137_{24}^{48}$  & $137_{24}^{49}$  & $163_{22}^{44}$  & $103_{17}^{34}$ & 0 &NSC  & \\ 
S43     &$ 13^{h} 25^{m} 28.20^{s} $&$ -43^{\circ} 02\arcmin 53\farcs4 $ &  $156_{21}^{43}$  & $108_{20}^{40}$  & $112_{20}^{41}$  & $98_{19}^{39}$  & $179_{22}^{45}$  & $170_{22}^{44}$ & 6 \\ 
S44     &$ 13^{h} 25^{m} 20.06^{s} $&$ -43^{\circ} 03\arcmin 10\farcs1 $ &  $166_{22}^{43}$  & $94_{21}^{43}$  & $134_{23}^{47}$  & $124_{23}^{46}$  & $146_{21}^{41}$  & $136_{20}^{39}$ & 2 &NSC & GC0587\\ 
S45     &$ 13^{h} 25^{m} 39.06^{s} $&$ -42^{\circ} 56\arcmin 53\farcs7 $ &  $78_{18}^{36}$  & $104_{17}^{35}$  & $51_{13}^{26}$  & $158_{21}^{42}$  & $87_{17}^{33}$  & $125_{23}^{45}$ & 2 & NSC \\ 
S46     &$ 13^{h} 25^{m} 25.15^{s} $&$ -43^{\circ} 01\arcmin 26\farcs9 $ &  $146_{24}^{49}$  & SC  & SC  & SC  & $125_{23}^{46}$  & $88_{21}^{41}$ & 0 & NSC & D \\ 
S47     &$ 13^{h} 25^{m} 23.07^{s} $&$ -43^{\circ} 01\arcmin 45\farcs6 $ &  $139_{21}^{43}$  & SC  & SC  & SC  & $128_{21}^{42}$  & $147_{22}^{44}$ & 2 & NSC \\ 
S48     &$ 13^{h} 25^{m} 32.42^{s} $&$ -42^{\circ} 58\arcmin 50\farcs2 $ &  $87_{17}^{34}$  & $143_{21}^{42}$  & $104_{18}^{36}$  & $133_{19}^{39}$  & $87_{16}^{32}$  & $121_{20}^{41}$ & 1 & $\beta $ NSC & GC0230 \\ 
S49     &$ 13^{h} 25^{m} 52.72^{s} $&$ -43^{\circ} 05\arcmin 46\farcs4 $ &  $72_{17}^{33}$  & $61_{15}^{30}$  & $108_{19}^{37}$  & $108_{20}^{40}$  & $71_{15}^{31}$  & $138_{21}^{42}$ & 9 & & GC0320 \\ 
S50     &$ 13^{h} 25^{m} 10.09^{s} $&$ -42^{\circ} 56\arcmin 08\farcs3 $ &  FoV  & FoV  & FoV & $138_{22}^{45}$  & $124_{21}^{43}$  & $75_{18}^{37}$ & 0 & FG \\ 
S51     &$ 13^{h} 25^{m} 45.47^{s} $&$ -42^{\circ} 58\arcmin 15\farcs8 $ &  $117_{20}^{40}$  & $114_{18}^{36}$  & $118_{18}^{36}$  & $123_{19}^{37}$  & $133_{20}^{40}$  & $95_{20}^{40}$ & 6 \\ 
S52     &$ 13^{h} 25^{m} 25.50^{s} $&$ -43^{\circ} 01\arcmin 29\farcs9 $ &  $127_{24}^{48}$  & SC  & SC & SC  & $0_{NA}^{NA}$  & $8_{8}^{22}$ & 0 & & N \\ 
S53     &$ 13^{h} 25^{m} 23.50^{s} $&$ -42^{\circ} 56\arcmin 51\farcs7 $ &  $111_{18}^{37}$  & $119_{20}^{41}$  & $131_{21}^{42}$  & $113_{18}^{37}$  & $95_{18}^{36}$  & $129_{22}^{44}$ & 1 & NSC  & \\ 
S54     &$ 13^{h} 25^{m} 23.63^{s} $&$ -43^{\circ} 03\arcmin 25\farcs7 $ &  $58_{13}^{26}$  & $58_{18}^{35}$  & $66_{18}^{37}$  & $82_{20}^{40}$  & $117_{18}^{37}$  & $25_{8}^{17}$ & 1 \\ 
S55     &$ 13^{h} 25^{m} 39.87^{s} $&$ -43^{\circ} 05\arcmin 01\farcs8 $ &  $110_{18}^{36}$  & $111_{19}^{38}$  & $103_{18}^{37}$  & $97_{19}^{39}$  & $92_{16}^{32}$  & $87_{16}^{32}$ & 0 & NSC & GC0266 \\ 
S56     &$ 13^{h} 25^{m} 38.10^{s} $&$ -43^{\circ} 05\arcmin 13\farcs7 $ &  $62_{14}^{27}$  & $71_{16}^{32}$  & $80_{16}^{33}$  & $74_{18}^{36}$  & $81_{15}^{31}$  & $112_{18}^{36}$ & 0 \\ 
S57     &$ 13^{h} 25^{m} 55.11^{s} $&$ -43^{\circ} 01\arcmin 18\farcs3 $ &  $96_{18}^{37}$  & $96_{16}^{33}$  & $77_{15}^{30}$  & $84_{16}^{31}$  & $91_{16}^{33}$  & $112_{20}^{41}$ & 0 \\ 
S58     &$ 13^{h} 25^{m} 24.76^{s} $&$ -43^{\circ} 01\arcmin 24\farcs6 $ &  $22_{15}^{31}$~R  & R  & R  & SC  & $77_{20}^{39}$  & $112_{22}^{43}$ & 0  &  & \\ 
S59     &$ 13^{h} 25^{m} 07.71^{s} $&$ -42^{\circ} 56\arcmin 29\farcs8 $ &  FoV  & FoV  & FoV  & $66_{18}^{35}$  & $103_{20}^{41}$  & $109_{20}^{40}$ & 2 & NSC & \\ 
S60     &$ 13^{h} 25^{m} 23.07^{s} $&$ -43^{\circ} 01\arcmin 34\farcs3 $ &  $103_{19}^{39}$~R  & SC  & SC  & SC  & $100_{19}^{39}$  & $80_{18}^{36}$ & 0 \\ 
S61     &$ 13^{h} 25^{m} 11.55^{s} $&$ -43^{\circ} 02\arcmin 26\farcs3 $ &  $103_{17}^{34}$  & SC  & SC  & SC  & $78_{15}^{31}$  & $109_{17}^{35}$ & 6 \\ 
\enddata
\tablecomments{Type column is explained in section~\ref{sec:select}, see also the extended discussion in section~\ref{s:disc}. N: Within $20\arcsec$ of the Cen A nucleus, R: Source is coincident with read-out streak, C: Source is coincident with chip-edge, FoV: Source is outside of the field-of-view, SC: Source confused,  GC: Coincident with globular cluster, D: Coincident with dust lane}
\end{deluxetable}
\pagebreak
\clearpage



%% file: thermal_results.tex
\begin{deluxetable}{lccccc}
\tabletypesize{\footnotesize}
\tablewidth{0pt}
\tablecaption{Spectral fits using inferred dominant model (0.5-8.0 keV) \label{tab:res} }
\tablehead{
\colhead{Source} &
\colhead{$N_{H}$} &
\colhead{${ kT_{in} }$} &
\colhead{${L_x~(0.5-10.0~{\rm keV}) }$} &
\colhead{${\rm \chi^2/dof }$} &
\colhead{Note}
\\
\colhead{} &
\colhead{${ \rm 10^{22}~cm^{-2}}$} &
\colhead{${\rm keV }$} &
\colhead{$ { \rm10^{37}~erg~s^{-1}}$} &
\colhead{} &
\colhead{}
}
\tablecolumns{6}
\startdata
\textbf{DBB} & & & & \\  \\ [3pt]
S3 & $0.08_{0.08}^{0.02}$ & $1.82_{0.13}^{0.14}$ & $24.58_{1.72}^{1.75},39.27_{2.36}^{2.41}$ & $120.7/140 $ \\ [3pt]
S4 & $0.65_{0.06}^{0.07}$ & $0.64_{0.03}^{0.03}$ & $21.06_{2.04}^{2.23},37.49_{3.09}^{3.41}$ & $206.9/203 $ & D \\ [3pt]
S4 & $0.59_{0.35}^{0.48}$ & $0.40_{0.11}^{0.13}$ & $6.15_{2.57}^{7.79}$ & $10.0/14 $ & D \\ [3pt]
S7 & $0.51_{0.06}^{0.07}$ & $1.49_{0.1}^{0.11}$ & $18.6_{1.4}^{1.5},27.2_{1.8}^{1.8}$ & $271.5/256 $ & D  \\ [3pt]
S8 & $0.08_{0.05}^{0.05}$ & $1.93_{0.17}^{0.25}$ & $15.71_{1.34}^{1.41},21.91_{1.76}^{1.85}$ & $135.7/132 $ & \\ [3pt]
S8 & $0.08_{0.08}^{0.03}$ & $1.1.74_{0.23}^{0.3}$ & $14.19_{1.29}^{1.37},21.07_{1.96}^{2.12}$ & $87.8/87 $ & \\ [3pt]
S13 & $0.40_{0.12}^{0.13}$ & $0.83_{0.09}^{0.1}$ & $10.57_{1.33}^{1.53},14.41_{1.69}^{1.93}$ & $110.7/89 $ & D \\ [3pt]
S13 & $0.39_{0.29}^{0.37}$ & $0.56_{0.14}^{0.17}$ & $4.34_{1.26}^{2.55},6.33_{1.77}^{3.72}$ & $37.5/30 $ & D \\ [3pt]
S13 & $0.35_{0.17}^{0.21}$ & $0.71_{0.11}^{0.13}$ & $11.65_{1.97}^{2.68}$ & $44.9/33 $ & D \\ [3pt]
S14 & $0.37_{0.11}^{0.12}$ & $0.61_{0.06}^{0.07}$ & $7.00_{1.03}^{1.24},12.33_{1.68}^{2.07}$ & $72.0/71 $ & D \\ [3pt]
S14 & $0.81_{0.23}^{0.28}$ & $0.52_{0.08}^{0.08}$ & $18.8_{4.4}^{7.4}$ & $14.6/27 $ & D \\ [3pt]
S19 & $0.09_{0.09}^{0.08}$ & $1.50_{0.17}^{0.13}$ & $5.45_{0.89}^{0.92},7.59_{0.8}^{0.85}$ & $119.3/135 $ \\ [3pt]
S22 & $0.08_{0.08}^{0.03}$ & $1.47_{0.14}^{0.16}$ & $4.93_{0.68}^{0.71},6.49_{0.73}^{0.76}$ & $63.6/67 $ \\ [3pt]
S26 & $0.08_{0.08}^{0.05}$ & $1.70_{0.19}^{0.24}$ & $6.13_{0.77}^{0.83},7.02_{0.89}^{0.97}$ & $51.6/60 $ & GC \\ [3pt]
S28 & $1.10_{0.4}^{0.45}$ & $2.09_{0.5}^{0.97}$ & $11.62_{1.66}^{1.76},11.9_{1.1}^{1.7}$ & $34.8/28 $ & D \\ [3pt]
S31 & $0.08_{0.08}^{0.05}$ & $1.08_{0.13}^{0.15}$ & $4.01_{0.53}^{0.57},4.2_{0.53}^{0.55}$ & $39.7/43 $ & \\ [3pt]
S31 & $0.12_{0.12}^{0.38}$ & $0.98_{0.31}^{0.27}$ & $3.1_{0.64}^{0.94}$ & $13.3/12 $  & \\ [3pt]
S32 & $0.08_{0.08}^{0.07}$ & $1.66_{0.23}^{0.27}$ & $5.26_{0.82}^{0.9},6.56_{0.98}^{1.08}$ & $28.2/50 $ & GC \\ [3pt]
S32 & $0.08_{0.08}^{0.13}$ & $1.65_{0.34}^{0.55}$ & $5.46_{0.97}^{1.19},7.67_{1.62}^{2.04}$ & $16.0/16 $  & GC\\ [3pt]
S34 & $0.08_{0.08}^{0.12}$ & $1.99_{0.37}^{0.37}$ & $4.71_{0.76}^{0.82},6.19_{0.92}^{1.01}$ & $52.6/67 $ \\ [3pt]
S38 & $0.60_{0.43}^{0.52}$ & $0.65_{0.17}^{0.25}$ & $5.48_{1.70}^{3.68}$ & $7.8/9 $  & D \\ [3pt]
S44 & $0.08_{0.08}^{0.08}$ & $1.21_{0.15}^{0.17}$ & $2.41_{0.4}^{0.42},2.93_{0.45}^{0.48}$ & $29.3/42 $ & GC \\ [3pt]
S45 & $0.18_{0.18}^{0.26}$ & $1.65_{0.41}^{0.63}$ & $2.86_{0.6}^{0.65},3.65_{0.7}^{0.78}$ & $28.5/21 $ &\\ [3pt]
\hline
 &  &  &  &   & \\ [0.5pt]
\textbf{PO} & & $\Gamma$ & & \\  \\ [3pt]
\hline
S3 & $0.13_{0.13}^{0.08}$ & $1.25_{0.11}^{0.13}$ & $36.29_{2.3}^{2.35},38.22_{2.65}^{2.74}$ & $109.6/117$ & \\ [3pt]
S5  & $0.11_{0.11}^{0.05}$ & $1.19_{0.08}^{0.09}$ & $29.37_{1.87}^{1.91},36.56_{2.25}^{2.3}$ & $180/214$ & GC\\ [3pt]
S8  & $0.08_{0.06}^{0.16}$ & $1.31_{0.2}^{0.28}$ & $21.11_{2.37}^{2.52}$ & $23.2/34$ & \\ [3pt]
S11  & $0.08_{0.08}^{0.17}$ & $1.29_{0.17}^{0.29}$ & $13.52_{1.75}^{1.87}$ & $17.3/23$ & GC\\ [3pt]
S11 & $0.09_{0.09}^{0.11}$ & $1.54_{0.08}^{0.19}$ & $12.44_{1.17}^{1.37},16.6_{1.36}^{1.54}$ & $72.3/73$  & GC\\ [3pt]
S16 & $0.14_{0.14}^{0.08}$ & $1.59_{0.14}^{0.15}$ & $10.05_{1.06}^{1.09},14.26_{1.58}^{1.62}$ & $126.5/129$ & GC \\ [3pt]
S17 & $0.16_{0.16}^{0.11}$ & $1.45_{0.16}^{0.17}$ & $10.77_{1.13}^{1.16},13.39_{3}^{3.07}$ & $93.1/105$& \\ [3pt]
S20  & $0.17_{0.17}^{0.10}$ & $1.74_{0.17}^{0.18}$ & $8.43_{0.95}^{0.98},11.25_{1.24}^{1.3}$ & $83.5/113$& \\ [3pt]
S22 & $0.08_{0.08}^{0.14}$ & $1.98_{0.23}^{0.35}$ & $5.85_{0.75}^{0.86}$ & $13.4/14$& \\ [3pt]
S25  & $0.11_{0.11}^{0.21}$ & $2.11_{0.21}^{0.47}$ & $6.24_{0.73}^{1.66}$ & $14/15$& \\ [3pt]
S29 & $0.16_{0.16}^{0.12}$ & $1.85_{0.21}^{0.24}$ & $5.86_{0.76}^{0.81},6.65_{0.82}^{0.87}$ & $65.1/64$& \\ [3pt]
S36 & $0.20_{0.2}^{0.23}$ & $2.39_{0.41}^{0.61}$ & $4.96_{1.03}^{2.11},5.08_{1.03}^{2.11}$ & $23.8/23$& \\ [3pt]
S42 & $0.13_{0.13}^{0.22}$ & $1.69_{0.30}^{0.36}$ & $3.97_{0.79}^{0.87},5.06_{0.99}^{1.06}$ & $46.5/44$ & \\ [3pt]
S46 & $0.11_{0.11}^{0.45}$ & $1.61_{0.21}^{0.61}$ & $3.00_{0.77}^{0.99},4.28_{0.83}^{1.25}$ & $28.2/27$ & D  \\ [3pt]
S47 & $0.11_{0.11}^{0.26}$ & $1.86_{0.2}^{0.53}$ & $3.26_{0.6}^{0.82},3.52_{0.65}^{0.79}$ & $14.7/23$& \\ [3pt]
S53 & $0.30_{0.30}^{0.32}$ & $2.29_{0.46}^{0.57}$ & $3.25_{0.86}^{1.66},3.91_{0.93}^{1.81}$ & $16/19$& \\ [3pt]
S55 & $0.19_{0.19}^{0.59}$ & $1.72_{0.31}^{0.83}$ & $3.34_{0.73}^{1.56},4.36_{0.99}^{2.05}$ & $15.1/14$ & GC \\ [3pt]
S59 & $0.11_{0.11}^{0.42}$ & $1.57_{0.26}^{0.67}$ & $2.91_{0.67}^{0.74},3.14_{0.67}^{0.91}$ & $9.8/12$& \\ [3pt]
\enddata
\tablecomments{Spectral fitting results with the appropriate simple model for thermally dominant (\textbf{DBB}) and power law dominant (\textbf{PO}) states, with $N_H$ forced to be above the Galactic value to reach a better estimate of the source luminosity.  GC denotes that the source is coincident with globular cluster while D indicates that a source is in the vicinity of the dust lane.}
\end{deluxetable}

%% file: tabtest.tex
\begin{landscape}
\begin{deluxetable}{lcccccccccc}
\tabletypesize{\footnotesize}
\tablewidth{0pt}
\tablecaption{Cen A Sources: Spectral fitting and State Identification \label{tab:tab2}}
\tablehead{
\colhead{Source} &
\colhead{ObsIDs} &
\colhead{${ N_{H}^{\rm PO}}$} &
\colhead{${ \Gamma^{\rm PO} }$} &
\colhead{${L_x~(0.5-10.0~{\rm keV})^{\rm PO} }$} &
\colhead{${\rm \chi^2/dof }$} &
\colhead{${N_{H}^{\rm DBB}}$} &
\colhead{${ kT_{in}^{\rm DBB} }$} &
\colhead{${L_x~(0.5-10.0~{\rm keV})^{\rm DBB} }$} &
\colhead{${\rm \chi^2/dof }$} &
\colhead{State} \\
\colhead{} &
\colhead{} &
\colhead{${ \rm 10^{22}~cm^{-2}}$} &
\colhead{} &
\colhead{$ { \rm 10^{37}~erg~s^{-1}}$} &
\colhead{} &
\colhead{${ \rm 10^{22}~cm^{-2}}$} &
\colhead{${\rm keV }$} &
\colhead{$ { \rm10^{37}~erg~s^{-1}}$} &
\colhead{} &
\colhead{}
}

\tablecolumns{11}
\startdata
S3 & $7797,7800$ & $0.13_{0.07}^{0.08}$ & $1.25_{0.12}^{0.13}$ & $36.2_{2.3}^{2.4},38.2_{2.6}^{2.7}$ & $109.6/117$ & $0.00_{0.00}^{0.05}$ & $2.61_{0.35}^{0.32}$ & $30.3_{2.3}^{2.5},31.7_{2.7}^{2.9}$ & $106.7/117 $ & P1  \\ [3pt]
S3 & $8489,8490$ & $0.21_{0.06}^{0.07}$ & $1.50_{0.11}^{0.11}$ & $31.7_{2.0}^{2.1},50.7_{2.7}^{2.8}$ & $131/140$ & $0.04_{0.04}^{0.05}$ & $1.92_{0.17}^{0.21}$ & $24.5_{1.7}^{1.8},39.2_{2.4}^{2.5}$ & $118.6/140 $ & T1  \\ [3pt]
S4 & $7797,7798,7799,7800$ & $1.25_{0.1}^{0.11}$ & $3.76_{0.18}^{0.19}$ & $93.77_{18.44}^{24.83},166.21_{31.51}^{42.48}$ & $241.2/203$ & $0.65_{0.06}^{0.07}$ & $0.64_{0.03}^{0.03}$ & $21.06_{2.04}^{2.23},37.49_{3.09}^{3.41}$ & $206.9/203 $ & I \\ [3pt]
S4 & $8489$ & $1.22_{0.51}^{0.7}$ & $5.07_{1.22}^{1.69}$ & $62_{46}^{447}$ & $10.5/14$ & $0.59_{0.35}^{0.48}$ & $0.40_{0.11}^{0.13}$ & $6.15_{2.57}^{7.79}$ & $10.0/14 $ & I \\ [3pt]
S5 & $7797,7800,8489,8490$ & $0.11_{0.05}^{0.05}$ & $1.19_{0.09}^{0.09}$ & $29.4_{1.9}^{1.9},36.6_{2.2}^{2.3}$ & $180/214$ & $0.00_{-0.82}^{0.00}$ & $2.83_{0.25}^{0.31}$ & $24.9_{1.8}^{1.9},31.1_{2.2}^{2.4}$ & $185.7/214 $ & P1,N  \\ [3pt]
S7 & $7797,7798,7799,8490,7800,8489$ & $0.89_{0.1}^{0.1}$ & $1.99_{0.11}^{0.12}$ & $28.8_{2.6}^{2.8},42.0_{3.4}^{3.8}$ & $275.7/256$ & $0.51_{0.06}^{0.07}$ & $1.49_{0.1}^{0.11}$ & $18.6_{1.4}^{1.5},27.2_{1.8}^{1.8}$ & $271.5/256 $ & I  \\ [3pt]
S8 & $7797,7798,8490$ & $0.29_{0.08}^{0.08}$ & $1.58_{0.13}^{0.14}$ & $20.6_{1.62}^{1.66},28.44_{2.07}^{2.14}$ & $128/132$ & $0.08_{0.05}^{0.05}$ & $1.93_{0.21}^{0.25}$ & $15.71_{1.34}^{1.41},21.91_{1.76}^{1.85}$ & $135.7/132 $ & T1 \\ [3pt]
S8 & $7799$ & $0.06_{0.06}^{0.16}$ & $1.31_{0.2}^{0.28}$ & $21.11_{2.37}^{2.52}$ & $23.2/34$ & $0.00_{-0.82}^{0}$ & $2.07_{0.38}^{0.57}$ & $16.8_{2.4}^{2.8}$ & $27.6/34 $ & P1,N \\ [3pt]
S8 & $7800,8489$ & $0.23_{0.09}^{0.10}$ & $1.57_{0.16}^{0.17}$ & $18.6_{1.51}^{1.56},27.58_{2.20}^{2.26}$ & $87.6/87$ & $0.04_{0.04}^{0.06}$ & $1.87_{0.23}^{0.3}$ & $14.19_{1.29}^{1.37},21.07_{1.96}^{2.12}$ & $87.8/87 $ & T1 \\ [3pt]
S11 & $7797$ & $0.07_{0.07}^{0.18}$ & $1.26_{0.23}^{0.32}$ & $13.5_{1.8}^{1.9}$ & $17.3/23$ & $0.00_{-0.82}^{0.00}$ & $2.19_{0.45}^{0.74}$ & $10.8_{1.7}^{2.1}$ & $17.8/23 $ & P1,N  \\ [3pt]
S11 & $7800,8489,8490$ & $0.09_{0.09}^{0.11}$ & $1.54_{0.18}^{0.19}$ & $12.4_{1.3}^{1.4},16.6_{1.5}^{1.5}$ & $72.3/73$ & $0.00_{-0.82}^{0.00}$ & $1.66_{0.17}^{0.20}$ & $9.52_{1.07}^{1.15},12.5_{1.2}^{1.4}$ & $83/73 $ & P1,N  \\ [3pt]
S12 & $7797,7798,7799,7800,8489,8490$ & $0.72_{0.19}^{0.21}$ & $1.12_{0.17}^{0.18}$ & $16.35_{1.85}^{1.88},26.54_{2.29}^{2.34}$ & $150.8/157$ & $0.49_{0.12}^{0.13}$ & $3.96_{0.87}^{1.73}$ & $14.49_{1.79}^{1.91},23.5_{2.3}^{2.5}$ & $153.6/157 $ & I  \\ [3pt]
S13 & $S7797,7798,8490$ & $0.9_{0.18}^{0.2}$ & $2.97_{0.29}^{0.31}$ & $27.05_{6.46}^{10.35},36.85_{8.44}^{13.55}$ & $122/89$ & $0.4_{0.12}^{0.13}$ & $0.83_{0.09}^{0.1}$ & $10.57_{1.33}^{1.53},14.41_{1.69}^{1.93}$ & $110.7/89 $ & I \\ [3pt]
S13 & $7800,8489$ & $1.02_{0.44}^{0.57}$ & $3.99_{0.89}^{1.16}$ & $22.53_{13.54}^{67.86},33_{19.79}^{100.31}$ & $41.3/30$ & $0.39_{0.29}^{0.37}$ & $0.56_{0.14}^{0.17}$ & $4.34_{1.26}^{2.55},6.33_{1.77}^{3.72}$ & $37.5/30 $ & I \\ [3pt]
S13 & $7799$ & $0.85_{0.27}^{0.34}$ & $3.33_{0.51}^{0.61}$ & $35.79_{13.48}^{33.25}$ & $58.2/33$ & $0.35_{0.17}^{0.21}$ & $0.71_{0.11}^{0.13}$ & $11.65_{1.97}^{2.68}$ & $44.9/33 $ & I \\ [3pt]
S14 & $7797,7798,7800$ & $0.95_{0.17}^{0.19}$ & $3.81_{0.35}^{0.39}$ & $31.4_{10.0}^{17.7},55.4_{17.4}^{30.7}$ & $78.1/71$ & $0.37_{0.11}^{0.12}$ & $0.61_{0.06}^{0.07}$ & $7.00_{1.03}^{1.24},12.33_{1.68}^{2.07}$ & $72/71 $ & I  \\ [3pt]
S14 & $7799$ & $1.65_{0.37}^{0.46}$ & $4.66_{0.64}^{0.78}$ & $192_{108}^{356}$ & $19.9/27$ & $0.81_{0.23}^{0.28}$ & $0.52_{0.08}^{0.08}$ & $18.8_{4.4}^{7.4}$ & $14.6/27 $ & I  \\ [3pt]
S16 & $7797,7798,7799,7800,8489,8490$ & $0.14_{0.08}^{0.08}$ & $1.59_{0.14}^{0.15}$ & $10.0_{1.1}^{1.1},14.3_{1.6}^{1.6}$ & $126.5/129$ & $0.00_{-0.82}^{0.00}$ & $1.67_{0.14}^{0.16}$ & $7.50_{0.83}^{0.86},10.5_{1.2}^{1.3}$ & $133.7/129 $ & P1,N  \\ [3pt]
S17 & $7797,7798,7799,7800,8489,8490$ & $0.16_{0.10}^{0.11}$ & $1.45_{0.16}^{0.17}$ & $10.77_{1.13}^{1.16},13.4_{3.0}^{3.1}$ & $93.1/105$ & $0.00_{0.00}^{0.06}$ & $1.98_{0.17}^{0.24}$ & $8.21_{0.94}^{0.8},10.28_{2.25}^{2.26}$ & $92.5/105 $ & P1  \\ [3pt]
S19 & $7797,7798,7799,7800,8489,8490$ & $0.35_{0.11}^{0.13}$ & $1.84_{0.19}^{0.2}$ & $7.91_{1.31}^{1.37},11.08_{1.23}^{1.36}$ & $121/135$ & $0.09_{0.07}^{0.08}$ & $1.5_{0.17}^{0.21}$ & $5.45_{0.89}^{0.92},7.59_{0.82}^{0.85}$ & $119.3/135 $ & T1  \\ [3pt]
S20 & $7797,7798,7799,8489,8490,7800$ & $0.17_{0.09}^{0.1}$ & $1.74_{0.17}^{0.18}$ & $8.43_{0.95}^{0.98},11.25_{1.24}^{1.30}$ & $83.5/113$ & $0.00_{-0.82}^{0.00}$ & $1.45_{0.12}^{0.14}$ & $5.98_{0.69}^{0.72},7.94_{0.89}^{0.94}$ & $90.2/113 $ & P1,N  \\ [3pt]
S22 & $7797$ & $0.0_{0.0}^{0.2}$ & $1.82_{0.21}^{0.46}$ & $5.61_{0.78}^{1.00}$ & $12.9/14$ & $0.00_{-0.82}^{0.00}$ & $1.16_{0.23}^{0.28}$ & $4.21_{0.66}^{0.74}$ & $20.9/14 $ & P1,N  \\ [3pt]
S22 & $7798,7799,7800,8489,8490$ & $0.21_{0.12}^{0.13}$ & $1.7_{0.21}^{0.22}$ & $6.90_{0.90}^{0.93},8.83_{1}^{1.07}$ & $60.1/67$ & $0.00_{-0.82}^{0.00}$ & $1.62_{0.16}^{0.19}$ & $4.94_{0.69}^{0.74},6.40_{0.74}^{0.78}$ & $60/67 $ & T2,M3,I  \\ [3pt]
S24 & $7797$ & $0.36_{0.33}^{0.4}$ & $1.54_{0.47}^{0.52}$ & $10.68_{1.62}^{2.11}$ & $9.4/14$ & $0.13_{0.13}^{0.26}$ & $1.85_{0.52}^{1.12}$ & $7.80_{1.34}^{1.75}$ & $8.6/14 $ & NA \\ [3pt]
S24 & $7798,7799,7800,8489$ & $0.64_{0.25}^{0.28}$ & $1.72_{0.27}^{0.3}$ & $10.4_{1.5}^{1.9},12.8_{1.8}^{2.4}$ & $74.8/67$ & $0.33_{0.16}^{0.18}$ & $1.68_{0.27}^{0.38}$ & $7.31_{0.92}^{0.96},9.03_{1.15}^{1.21}$ & $67.2/67 $ & I  \\ [3pt]
S25 & $7797$ & $0.11_{0.11}^{0.21}$ & $2.11_{0.36}^{0.47}$ & $6.24_{1.08}^{1.66}$ & $14/15$ & $0.00_{-0.82}^{0.00}$ & $1_{0.15}^{0.18}$ & $4.28_{0.56}^{0.59}$ & $14.1/15 $ & P1,N  \\ [3pt]
S26 & $7797,7798,7799,7800$ & $0.22_{0.12}^{0.13}$ & $1.57_{0.20}^{0.21}$ & $8.29_{0.98}^{1},9.38_{1.13}^{1.17}$ & $49.1/60$ & $0.02_{0.02}^{0.09}$ & $1.84_{0.28}^{0.32}$ & $6.14_{0.85}^{0.92},7.03_{0.92}^{1}$ & $50.3/60 $ & T1  \\ [3pt]
S28 & $7797,8490$ & $0.51_{0.48}^{0.53}$ & $0.87_{0.39}^{0.41}$ & $1.20_{1.67}^{1.78},17.07_{2.38}^{2.54}$ & $17.5/27$ & $0.44_{0.30}^{0.36}$ & $5.60_{2.72}^{NA}$ & $11.1_{1.94}^{2.43},15.74_{2.80}^{3.54}$ & $17.0/31 $ & I  \\ [3pt]
S28 & $7800,8489$ & $1.66_{0.64}^{0.72}$ & $1.71_{0.45}^{0.49}$ & $16.58_{3.21}^{6.19},16.95_{3.3}^{6.58}$ & $36.7/28$ & $1.10_{0.4}^{0.45}$ & $2.09_{0.5}^{0.97}$ & $11.62_{1.66}^{1.76},11.9_{1.1}^{1.7}$ & $34.8/28 $ & I  \\ [3pt]
S29 & $7797,7799,7800,8489,8490$ & $0.16_{0.11}^{0.12}$ & $1.85_{0.23}^{0.24}$ & $5.86_{0.76}^{0.81},6.65_{0.82}^{0.87}$ & $65.1/64$ & $0.00_{-0.82}^{0.00}$ & $1.28_{0.13}^{0.15}$ & $4.00_{0.53}^{0.55},4.45_{0.58}^{0.62}$ & $80.2/64 $ & P1,N  \\ [3pt]
S29 & $7798$ & $0.83_{0.39}^{0.50}$ & $2.55_{0.55}^{0.65}$ & $10.30_{3.30}^{8.80}$ & $9/10$ & $0.38_{0.26}^{0.32}$ & $1.04_{0.22}^{0.32}$ & $5.14_{0.96}^{1.24}$ & $8/10 $ & I  \\ [3pt]
S31 & $7797,7800,8490$ & $0.26_{0.15}^{0.17}$ & $2.08_{0.31}^{0.34}$ & $6.23_{0.93}^{1.19},6.51_{0.95}^{1.25}$ & $30/43$ & $0.00_{0.00}^{0.09}$ & $1.20_{0.18}^{0.11}$ & $3.91_{0.45}^{0.59},4.09_{0.47}^{0.56}$ & $37.4/43 $ & T1  \\ [3pt]
S31 & $7798, 8489$ & $0.00_{0.00}^{0.82}$ & $1.29_{0.19}^{0.20}$ & $6.31_{1.37}^{1.80},6.69_{1.11}^{1.00}$ & $21.3/27$ & $0.00_{0.00}^{0.82}$ & $1.85_{0.40}^{0.66}$ & $4.72_{0.86}^{1.10},5.17_{1.00}^{1.35}$ & $24.5/27 $ & NA \\ [3pt]
S31 & $7799$ & $0.53_{0.43}^{0.60}$ & $2.58_{0.77}^{1.06}$ & $6.24_{2.22}^{9.34}$ & $13.7/12$ & $0.12_{0.12}^{0.38}$ & $0.98_{0.31}^{0.58}$ & $3.10_{0.74}^{0.94}$ & $13.3/12 $ & T1  \\ [3pt]
S32 & $7797,7798,7799,7800$ & $0.26_{0.16}^{0.18}$ & $1.61_{0.26}^{0.28}$ & $7.22_{1.09}^{1.16},8.98_{1.25}^{1.32}$ & $26.5/50$ & $0.03_{0.03}^{0.12}$ & $1.79_{0.32}^{0.36}$ & $5.27_{0.85}^{0.94},6.58_{1.01}^{1.13}$ & $27.5/50 $ & T1  \\ [3pt]
S32 & $8489,8490$ & $0.00_{-0.82}^{0.00}$ & $1.32_{0.22}^{0.23}$ & $7.18_{1.21}^{1.35},10.61_{2.1}^{2.39}$ & $13/16$ & $0.00_{-0.82}^{0.00}$ & $1.8_{0.40}^{0.69}$ & $5.45_{1.03}^{1.30},7.75_{1.71}^{2.21}$ & $14.3/16 $ & T2,M3,I,N,M2  \\ [3pt]
S34 & $7797,7798,7799,7800,8489$ & $0.32_{0.16}^{0.18}$ & $1.55_{0.24}^{0.26}$ & $6.23_{0.93}^{0.96},8.14_{1.08}^{1.13}$ & $48.2/67$ & $0.08_{0.08}^{0.12}$ & $2.00_{0.37}^{0.58}$ & $4.72_{0.76}^{0.84},6.19_{0.92}^{1.03}$ & $52.6/67 $ & T1  \\ [3pt]
S34 & $8490$ & $0.97_{0.43}^{0.55}$ & $2.26_{0.52}^{0.62}$ & $11.34_{3.03}^{7.61}$ & $8.5/13$ & $0.49_{0.27}^{0.34}$ & $1.32_{0.3}^{0.46}$ & $6.46_{1.11}^{1.23}$ & $7.3/13 $ & I  \\ [3pt]
S35 & $8490$ & $0.51_{0.49}^{0.6}$ & $1.81_{0.66}^{0.74}$ & $8.02_{1.54}^{4.0}$ & $7.1/10$ & $0.20_{0.20}^{0.40}$ & $1.49_{0.47}^{1.05}$ & $5.28_{0.98}^{1.28}$ & $6.9/10 $ & NA \\ [3pt]
S36 & $7797,7798,8490$ & $0.20_{0.18}^{0.21}$ & $1.54_{0.34}^{0.36}$ & $4.5_{0.86}^{0.94},7.31_{1.06}^{1.14}$ & $30.5/31$ & $0.02_{0.02}^{0.14}$ & $1.71_{0.38}^{0.46}$ & $3.20_{0.66}^{0.82},5.17_{0.87}^{1.09}$ & $29.7/31 $ & NA \\ [3pt]
S36 & $7799,7800$ & $0.20_{0.20}^{0.23}$ & $2.39_{0.50}^{0.61}$ & $4.96_{1.04}^{2.11},5.08_{1.04}^{2.11}$ & $23.8/23$ & $0.00_{-0.82}^{0.00}$ & $0.87_{0.15}^{0.19}$ & $2.85_{0.45}^{0.49},2.92_{0.45}^{0.50}$ & $33.4/23 $ & P1,N  \\ [3pt]
S38 & $8490$ & $1.41_{0.64}^{0.76}$ & $3.87_{1.00}^{1.25}$ & $31.34_{20.9}^{126.4}$ & $6.1/9$ & $0.60_{0.43}^{0.52}$ & $0.65_{0.17}^{0.25}$ & $5.48_{1.70}^{3.68}$ & $7.8/9 $ & I  \\ [3pt]
S39 & $7797,7798,7799$ & $0.93_{0.38}^{0.43}$ & $1.98_{0.44}^{0.48}$ & $7.05_{1.44}^{2.65},9.28_{1.9}^{3.42}$ & $37.2/30$ & $0.41_{0.24}^{0.27}$ & $1.65_{0.38}^{0.65}$ & $4.45_{0.73}^{0.79},5.80_{0.96}^{1.06}$ & $41.3/30 $ & I  \\ [3pt]
S39 & $7799,7800,8490$ & $0.92_{0.37}^{0.45}$ & $2.34_{0.47}^{0.53}$ & $8.11_{2.22}^{4.8},10.08_{2.71}^{5.64}$ & $19.6/26$ & $0.43_{0.24}^{0.29}$ & $1.2_{0.24}^{0.36}$ & $4.28_{0.79}^{0.90},5.46_{1.07}^{1.16}$ & $21.3/26 $ & I  \\ [3pt]
S42 & $7797,7799,7800,8489,8490$ & $0.13_{0.13}^{0.22}$ & $1.69_{0.30}^{0.36}$ & $3.97_{0.79}^{0.87},5.06_{0.99}^{1.06}$ & $46.5/44$ & $0.00_{-0.82}^{0.00}$ & $1.38_{0.19}^{0.25}$ & $2.78_{0.54}^{0.58},3.54_{0.71}^{0.78}$ & $44.7/44 $ & P1,N  \\ [3pt]
S44 & $7797,7799,7800,8489,8490$ & $0.30_{0.18}^{0.20}$ & $2.03_{0.31}^{0.34}$ & $3.71_{0.67}^{0.84},4.37_{0.81}^{1.05}$ & $30.2/42$ & $0.01_{0.01}^{0.13}$ & $1.30_{0.21}^{0.22}$ & $2.37_{0.40}^{0.42},2.83_{0.42}^{0.50}$ & $28.4/42 $ & T1  \\ [3pt]
S45 & $7798,7800,8490$ & $0.48_{0.32}^{0.39}$ & $1.76_{0.44}^{0.49}$ & $4.07_{0.91}^{1.37},5.19_{1.02}^{1.51}$ & $27.7/21$ & $0.18_{0.18}^{0.26}$ & $1.65_{0.41}^{0.77}$ & $2.86_{0.60}^{0.65},3.65_{0.70}^{0.78}$ & $28.5/21 $ & T1  \\ [3pt]
S46 & $7797,8489,8490$ & $0.11_{0.11}^{0.45}$ & $1.61_{0.32}^{0.61}$ & $3.00_{0.81}^{0.99},4.28_{0.95}^{1.25}$ & $28.2/27$ & $0.00_{0.00}^{0.21}$ & $1.51_{0.36}^{0.48}$ & $2.16_{0.57}^{0.70},3.13_{0.70}^{0.84}$ & $26.3/27 $ & P1  \\ [3pt]
S47 & $7797,8489,8490$ & $0.11_{0.11}^{0.26}$ & $1.86_{0.35}^{0.53}$ & $3.26_{0.63}^{0.82},3.52_{0.68}^{0.79}$ & $14.7/23$ & $0.00_{-0.82}^{0.00}$ & $1.13_{0.19}^{0.26}$ & $2.16_{0.40}^{0.45},2.32_{0.45}^{0.53}$ & $17.3/23 $ & P1,N  \\ [3pt]
S48 & $7798,8490,7799$ & $0.00_{-0.82}^{0.00}$ & $0.75_{0.22}^{0.22}$ & $5.17_{1.24}^{1.49},6.97_{1.31}^{1.45}$ & $12.8/18$ & $NA_{NA}^{NA}$ & $NA_{NA}^{NA}$ & $NA_{}^{},NA_{}^{}$ & $NA/NA $ & M2  \\ [3pt]
S53 & $7797,7800,8490$ & $0.30_{0.28}^{0.32}$ & $2.29_{0.49}^{0.57}$ & $3.25_{0.86}^{1.66},3.91_{0.93}^{1.81}$ & $16/19$ & $0.00_{-0.82}^{0.00}$ & $1.07_{0.17}^{0.21}$ & $1.93_{0.35}^{0.37},2.28_{0.43}^{0.45}$ & $17.8/19 $ & P1,N  \\ [3pt]
S55 & $7797,7798,7799$ & $0.19_{0.19}^{0.59}$ & $1.72_{0.45}^{0.83}$ & $3.34_{0.75}^{1.55},4.36_{1.01}^{2.05}$ & $15.1/14$ & $0.00_{-0.82}^{0.00}$ & $1.42_{0.32}^{0.53}$ & $2.27_{0.52}^{0.65},2.94_{0.68}^{0.9}$ & $17/14 $ & P1,N  \\ [3pt]
S59 & $8489,8490$ & $0.11_{0.11}^{0.42}$ & $1.57_{0.39}^{0.67}$ & $2.91_{0.68}^{0.74},3.14_{0.81}^{0.92}$ & $9.8/12$ & $0.00_{-0.82}^{0.00}$ & $1.55_{0.41}^{0.74}$ & $2.07_{0.52}^{0.67},2.3_{0.62}^{0.83}$ & $10.6/12 $ & P1,N  \\ [3pt]
\enddata
\tablecomments{Results from fitting absorbed power law ($PO$) and disk blackbody ($DBB$) models with all parameters free to vary and $95\%$ confidence intervals.  The state column denotes the spectral states that a given set of spectra are consistent with, based on the classification scheme proposed by \cite{2010ApJ...725.1805B}.  T indicates a thermally dominant state, P indicates power law dominant state, M suggests that there is significant, unmodelled emission from another component ($>30-40\%$), N suggests a result was not-produced by the \cite{2010ApJ...725.1805B} simulations and I indicates there is additional line-of-sight absorption, when both simple models have $N_H$ significantly above the Galactic value.  The extent to which a component dominates is indicated by the associated number, i.e. 1 implies the state is almost completely dominated by that component, while 2 indicates that there is some unmodelled component also present.  Instances of $N_H^{DBB}=0.00_{-0.82}^{0.00}$ for some fits are the result of not being able to constrain an upper-limit in Xspec.  We believe the source to be in the power law state at these times, with $N_H^{PO}$ consistent with the Galactic value.  In the state column, `NA' denotes where no inference can be made regarding the state (see section~\ref{sec:uncha}) and for S48 shows where no fit could be achieved (i.e. $\chi/dof > 2$), which is discussed in section~\ref{sec:S48}.} 
\tablenotetext{a}{Minimal and maximal values of $L_x$ are shown for absorbed power law (PO) and disk blackbody (DBB) models}

\end{deluxetable}
\clearpage
\end{landscape}